\documentclass[useAMS,usenatbib]{mnras}
\usepackage[dvips]{graphicx}
\usepackage[english]{babel}
\usepackage{amsmath}
\usepackage{amssymb}
\usepackage{textcomp}
\usepackage{natbib}
\usepackage{multirow}
\usepackage{bold-extra} 

\newcommand{\Mh}{{\rm M}_{\rm halo}}
\newcommand{\Msun}{{\rm M}_{\odot}}
\newcommand{\Mdotta}{\dot{\rm M}_{\rm ta}}
\newcommand{\Mdotstar}{\dot{\rm M}_{\star}}
\newcommand{\tcool}{t_{\rm cool}}
\newcommand{\tff}{t_{\rm ff}}
\newcommand{\rvir}{r_{\rm vir}}
\newcommand{\rta}{r_{\rm ta}}
\newcommand{\rsh}{r_{\rm sh}}
\newcommand{\rcore}{r_{\rm core}}
\newcommand{\rgal}{r_{\rm gal}}
\newcommand{\vw}{v_{\rm wind}}
\newcommand{\ve}{v_{\rm esc}}

\newcommand{\be}{\begin{equation}}
\newcommand{\ee}{\end{equation}}
\newcommand{\bea}{\begin{align}}
\newcommand{\eea}{\end{align}}
\newcommand{\beg}{\begin{gather}}
\newcommand{\eeg}{\end{gather}}

\title[Star formation feedback and the CGM]{The Impact of Star Formation Feedback on the Circumgalactic Medium}
\author[Fielding, Quataert, McCourt, \& Thompson]{
Drummond~Fielding$^{1}$\thanks{E-mail: dfielding@berkeley.edu}, 
Eliot~Quataert$^{1}$, 
Michael~McCourt$^{2}$, 
and Todd~A.~Thompson$^{3}$\\
 $^1$Astronomy Department and Theoretical Astrophysics Center, University of California Berkeley, Berkeley, CA 94720, USA\\
 $^2$Department of Physics, University of California Santa Barbara, Santa Barbara, CA 93106, USA\\
 $^3$Department of Astronomy and Center for Cosmology \& Astro-Particle Physics, The Ohio State University, Columbus, Ohio 43210}

\begin{document}

\date{Accepted 2016 December 19. Received 2016 November 28; in original form 2016 June 10}

\pagerange{\pageref{firstpage}--\pageref{lastpage}} \pubyear{2016}

\maketitle

\label{firstpage}

\begin{abstract}
We use idealized three-dimensional hydrodynamic simulations to study the dynamics and thermal structure of the circumgalactic medium (CGM). Our simulations quantify the role of cooling, stellar feedback driven galactic winds, and cosmological gas accretion in setting the properties of the CGM in dark matter haloes ranging from $10^{11}-10^{12}$\,M$_\odot$. Our simulations support a conceptual picture in which the properties of the CGM, and the key physics governing it, change markedly near a critical halo mass of ${\rm M}_{\rm crit} \approx 10^{11.5}~\Msun$. As in calculations without stellar feedback, above ${\rm M}_{\rm crit}$ halo gas is supported by thermal pressure created in the virial shock. The thermal properties at small radii are regulated by feedback triggered when $t_{\rm cool} / t_{\rm ff} \lesssim 10$ in the hot gas. Below ${\rm M}_{\rm crit}$, however, there is no thermally supported halo and self-regulation at $t_{\rm cool} / t_{\rm ff} \sim10$ does not apply. Instead, the gas is out of hydrostatic equilibrium and largely supported against gravity by bulk flows (turbulence and coherent inflow/outflow) arising from the interaction between cosmological gas inflow and outflowing galactic winds. In these lower mass halos the phase structure depends sensitively on the outflows' energy per unit mass and mass-loading, which may allow measurements of the CGM thermal state to constrain the nature of galactic winds. Our simulations account for some of the properties of the multiphase halo gas inferred from quasar absorption line observations, including the presence of significant mass at a wide range of temperatures, and the characteristic  O\thinspace\textsc{vi} and C\thinspace\textsc{iv} column densities and kinematics. However, we under-predict the neutral hydrogen content of the $z\sim0$ CGM.
\end{abstract}

\begin{keywords}
galaxies: haloes -- galaxies: formation -- galaxies: evolution -- intergalactic medium -- quasars: absorption lines -- cosmology: theory
\vspace*{0.25cm}   
\end{keywords}

\section{Introduction}
The classic paradigm in galaxy formation is that dark matter haloes are initially filled with hot gas heated to the virial temperature by an accretion shock. This virialized gas settles into rough hydrostatic equilibrium in the dark matter potential. In sufficiently massive haloes the virialized gas cannot cool on a Hubble time and the galactic accretion rate is set by the cooling rate of the halo gas. However, the shock heated gas cools quickly and rapidly loses pressure support in less massive haloes. The critical transition between rapid and slow cooling occurs at dark matter halo masses on the order of $\sim 10^{11.5}~\Msun$, relatively independent of redshift \citep{ReesOstriker+77,Silk77,Binney77}. 

Using analytic calculations and spherically symmetric simulations \cite{BirnboimDekel03} sharpened the understanding of galaxy growth in haloes below ${\sim} 10^{11.5}~\Msun$ by showing that inflowing gas does not form an accretion shock near the virial radius when the cooling time of the post-shock gas $t_{\rm cool}$ is less than the free fall time $t_{\rm ff}$. Their calculations, however, neglected feedback processes. 

Cosmological simulations have subsequently borne out these ideas with greater realism \citep[e.g.,][]{Keres+05, Dekel+09}. These simulations have verified the existence of a critical halo mass ${\sim}10^{11.5}~\Msun$ below which accretion proceeds via `cold streams' that penetrate directly to small radii. In more massive haloes the classic picture remains appropriate with gas shock heated to the virial temperature. The exact transition between cold and hot accretion remains somewhat uncertain, however, with different numerical techniques providing somewhat different answers \citep{Keres+12,Nelson+13}. 

The transition in how galaxies acquire their gas appears to be key for understanding many properties of galaxies. Notably, this critical halo mass corresponds to the stellar mass where galaxies transition from being predominantly blue and star forming to red and quiescent \citep[e.g.,][]{Yang+09}, and to roughly the peak in the stellar-to-halo mass ratio \citep[e.g.,][]{Behroozi+10}. 

In parallel to this improved understanding of halo and galaxy accretion, there has been rapid advancement in our understanding of the properties and dynamics of halo gas. For cluster and group mass haloes ($\geq10^{13}~\Msun$) thermal instability triggered feedback regulation \citep[e.g.,][]{McCourt+12,Sharma+12a,Sharma+12b,Li+15} has proved successful in explaining some of the properties of both the cool and hot intracluster (intragroup) medium \citep{Voit+15}. Simultaneously, quasar absorption line observations have begun to provide detailed quantitative constraints on the mass, metal content, and phase structure of gas in galaxy mass dark matter haloes \citep[e.g.,][]{Steidel+10, Tumlinson+11, Rudie+12, Werk+14, Borthakur+15}. 

In this paper, we adopt an idealized approach to studying the gaseous haloes of galaxies, the circumgalactic medium, i.e., CGM. The interplay of cooling, galactic winds driven by stellar feedback, and cosmological accretion of gas shape the CGM and determine its dynamics and thermal structure.  Our aim in this paper is in part to assess the impact of stellar feedback on what has become the established understanding of the dark matter halo mass dependence of virial shock stability. Furthermore, we seek to determine how the phase structure of halo gas changes with halo mass and feedback parameterization. These topics are, of course, also addressed by fully cosmological simulations focused on the CGM that incorporate stellar feedback \citep[e.g.,][]{vandeVoort+12,Shen+13,Ford+13, CAFG+15, CAFG+16,vandeVoort+16}. Cosmological simulations have also been used to perform controlled experiments in which the effect of different feedback models on the star formation rate, inflow and outflow rates, and the CGM structure and corresponding observables are studied \citep[e.g.,][]{FaucherGiguere+11, Oppenheimer+10,Nelson+15, Hummels+13, Suresh+15a,Marasco+15, Rahmati+15,Liang+16}. 
Here we adopt a complementary approach and use idealized three dimensional hydrodynamic simulations that sacrifice some degree of realism, but provide more control and better physical insight into the dominant processes.  

In this initial study we make several important simplifications. The most readily apparent relative to cosmological simulations is that we do not consider filamentary accretion and instead feed gas into our haloes quasi-spherically. This choice was made because of the computational subtleties in resolving instabilities between inflow filaments and halo gas \citep{Keres+12,Nelson+13,Lecoanet+15, Mandelker2016}, which is in some sense a distinct (albeit important) set of questions from those we address here.
 Additionally, we make the fairly standard simplification of solving the ideal hydrodynamics equations only. Magnetic fields, (anisotropic) conduction \citep{Balbus01,Quataert08,McCourt+11}, viscosity \citep{Kunz11,Parrish+12}, and cosmic rays \citep{Booth+13} may be important for properly modeling the CGM. In future studies we plan to relax these assumptions while maintaining the controlled and idealized nature of our simulations.

The structure of this paper is as follows. We describe our computational set-up in Section \ref{method}. In Section \ref{results} we present the results of our simulations, focusing on the halo mass dependence of the CGM properties, how the CGM changes as we modify the feedback physics, and a comparison of our results to observations of the $z\sim0$ CGM. In Section \ref{Discussion} we conclude with a summary of our results and discuss the implications and future directions of our work.

\section{Method} \label{method}
We study the long term evolution of gas in galactic haloes -- in particular, how the evolution changes with halo mass and with feedback efficiency/strength. The numerical experiment we designed models the relevant physical processes while remaining simple enough for us to readily determine what causes the resulting behavior. Our model for the galactic halo takes into account the gravitational potential of the dark matter, optically thin radiative cooling, ongoing cosmological accretion, and galactic feedback that is triggered when gas is accreted on to the central galaxy. We ran 3-dimensional hydrodynamic simulations with an ideal gas equation of state using the \textsc{athena} code \citep{Stone+08, Gardiner+08}, which integrates the standard fluid equations. We make use of the static mesh refinement capabilities of \textsc{athena} to reach high resolution in the central regions of the haloes.

At the scales we are interested in, dark matter dominates the gravitational potential, so we do not include any baryonic contribution to the gravitational potential in our calculations. We treat the dark matter as a static potential that follows an NFW profile \citep{NFW}. We adopt the common `200m' definition of the mass and $\rvir$ of the halo. They are defined such that the mean density of the halo is 200 times $\bar{\rho}_m$, the mean matter density of the Universe: $\Mh = \mathrm{M}_{200 m} = 200 \bar{\rho}_m (4 \pi / 3) \rvir^3$. We assume a $\Lambda$CDM cosmology with $(\Omega_m, \Omega_\Lambda, H_\circ)$ = (0.27, 0.73, 70 km/s/Mpc). 

We restrict our attention to the $z=0$ universe. However, our results are generally applicable to a wide range of redshifts because the dynamics are not expected to change much with redshift at fixed halo mass \citep{DekelBirnboim06}. This is due to the very weak redshift dependence of $\tcool/\tff$ at the accretion shock of a halo. We have confirmed this in our setup with a small set of simulations, but we leave a detailed investigation of the redshift dependence to a future work. 

In keeping with the idealized nature of these calculations we keep the metallicity of the gas fixed at one-third solar, including the cosmologically inflowing gas and the galactic wind gas, which are likely less and more metal enriched, respectively. 
{ All gas is assumed to be in ionization equilibrium with the photo-ionizing photons coming from the meta-galactic UV/X-ray background \citep{HaardtMadau01}, i.e., no local sources. The assumption of ionization equilibrium is likely valid at most times in our simulations because at the characteristic CGM densities and temperatures the cooling time is a few times shorter than the recombination time, $t_{\rm rec}$, for most of the relevant ionic species (however this may also depend on the degree of turbulence in the medium, see \citealt{Gray+16}). In particular, for O\thinspace\textsc{vi} $t_{\rm rec}\sim 3.5 t_{\rm cool}$ \citep{NaharPradhan03}. Furthermore, \cite{Oppenheimer+16} demonstrated that taking non-equilibrium effects into consideration does not significantly alter the cooling of halo gas nor the resulting O\thinspace\textsc{vi} column densities}.
{ We adopt the equilibrium cooling (and heating) rates tabulated by \cite{Wiersma+09}. The difference between these cooling rates and collisional ionization only cooling rates \citep[e.g.,][]{SD93} can be significant at the typical, low densities of the haloes we consider. }Additionally, we do not allow gas to cool below $T= 10^4$~K. 
This temperature floor is somewhat redundant given the low temperature photoionization heating, but ensures that unresolved dense clumps do not become under-pressurized and overly massive.

We include cosmological accretion of gas by feeding in cold gas at the turn-around radius, $r_{\rm ta} = 2 \rvir$, which is the outer boundary of our computational domain.  This accretion is quasi-spherical ($\delta \rho/\rho \sim 0.3$ isobaric, isotropic perturbations are introduced to break spherical symmetry, the details of the perturbations are discussed below); as we discuss in Section~\ref{Discussion}, other accretion geometries, such as filaments, will be considered in future work.  
The accretion rate at the turn around radius, $\Mdotta$, is calibrated to match the mean rates measured by \cite{Mcbride+09} in the large, dark matter only, Millennium Simulation \citep{Millennium} -- scaled appropriately by the cosmic baryon fraction $f_{\rm b} = 0.17$. 
Explicitly, we use $\dot{\rm M}_{\rm ta}~=~7 \, \mathrm{M}_\odot \, \mathrm{yr}^{-1} (\Mh/10^{12} \Msun)$. In practice, the desired accretion rate is achieved by resetting the density and velocity at $r_{\rm ta}$ every time step. The velocity at $r_{\rm ta}$ is set to be $v_{\rm ta} = 0.1 v_{\rm vir} = 0.1 ( G \Mh\rvir^{-1})^{1/2}$, and the density is $\rho_{\rm ta} = \dot{\rm M}_{\rm ta} (4 \pi r_{\rm ta}^2 v_{\rm ta})^{-1}$. The results are insensitive to the exact value of $v_{\rm ta}$ as long as the velocity is small.
 
We use a physically motivated mechanical galactic feedback model that depends on the wind velocity $\vw$ and the mass loading factor of the wind $\eta$, which is defined such that 
\be
\dot{\rm M}_\star\eta = \dot{\rm M}_{\rm out} , ~\dot{{\rm M}}_{\rm in} \frac{\eta}{\eta+1}  = \dot{{\rm M}}_{\rm out} \label{eq:eta}. 
\ee
We do not simulate star formation so the star formation rate in equation (\ref{eq:eta}) instead represents the rate at which gas is excised from the inner edge of the domain.

\begin{table}
\begin{center}
\caption{Simulation parameters}
\label{Feedback parameters}
\hspace*{-0.5cm}   
\begin{tabular}{|l@{\hskip -0.03in}|c@{\hskip -0.03in}|c@{\hskip -0.03in}|c@{\hskip -0.03in}|c@{\hskip -0.01in}|c@{\hskip -0.02in}|c@{\hskip -0.02in}|c@{\hskip -0.0in}|}
\hline \hline
\multirow{2}{*}{$\Mh$}$^{a}$ & \multirow{2}{*}{$r_{\rm vir}$}$^{b}$ & \multirow{2}{*}{$v_{\rm vir}$}$^{c}$ & \multirow{2}{*}{$\dot{\rm M}_{\rm ta}$}$^{d}$ & \multirow{2}{*}{$\eta$} & \multirow{2}{*}{$\left(\frac{\vw}{\ve}\right)^2$} & $\epsilon_\star$ & \multirow{2}{*}{label}          \\
&&&&&& {\footnotesize($\times10^{-6}$)} & \\ \hline\hline

\multirow{4}{*}{$10^{11}$}   & \multirow{4}{*}{148} & \multirow{4}{*}{54} & \multirow{4}{*}{0.7} & 5  & 1  & $1.0$ & fiducial high $\eta$ \\
                             &                      &                     &                      & 5  & 3  & $3.0$ & strong high $\eta$ \\
                             &                      &                     &                      & 0.3& 4.5& $0.3$ & fiducial low $\eta$ \\ 
                             &                      &                     &                      & 0.3& 9  & $0.6$ & strong low $\eta$ \\  \hline
\multirow{2}{*}{$10^{11.5}$} & \multirow{2}{*}{217}  & \multirow{2}{*}{79} & \multirow{2}{*}{2.2} & 3   & 2    & $2.6$  & fiducial high $\eta$ \\   
                             &                       &                     &                      & 0.3 & 6.75 & $0.9$  & fiducial low $\eta$ \\    \hline
\multirow{2}{*}{$10^{12}$}   & \multirow{2}{*}{319}  & \multirow{2}{*}{116} & \multirow{2}{*}{7.0} & 2    & 3 & $5.4$& fiducial high $\eta$ \\ 
                             &                       &                      &                      & 0.3  & 9 & $2.4$& fiducial low $\eta$ \\  \hline\hline
\end{tabular}
\end{center}
{\footnotesize $^{a}$ in units of $\Msun$; $^{b}$ in units of kpc; $^{c}$ in units of km s$^{-1}$; $^{d}$ in units of $\mathrm{M}_\odot ~ \mathrm{yr}^{-1}$. For each halo mass, we ran simulations with both high and low mass-loading $\eta$ and corresponding lower or higher $\vw$, respectively. For each choice of $\eta$ for the $10^{11}~\Msun$ halo we adopted a fiducial (smaller) and a `strong' (larger) $\vw$. Note that $\ve\approx3.5 \,v_{\rm vir}$ at $\rgal$, where the wind is launched, which can be used to determine $\vw$. We make use of static mesh refinement in these simulations to increase the resolution in the centers of haloes. The fiducial spatial resolution is 57 cells per $\rvir$ ($\Delta x = 5.6$~kpc~$\mathrm{M}_{12}^{1/3}$, where $\mathrm{M}_{12}=\Mh (10^{12} \Msun)^{-1}$), 114 cells per $\rvir$ ($\Delta x = 2.8$~kpc~$\mathrm{M}_{12}^{1/3}$), and 228 cells per $\rvir$ ($\Delta x = 1.4$~kpc~$\mathrm{M}_{12}^{1/3}$), for $r>1.125\rvir$, $1.125\rvir>r>0.5625\rvir$, and $r<0.5625\rvir$, respectively. In Appendix \ref{Appendix} we sensitivity of our results on spatial resolution and find our results are well converged. }
\end{table}

We are not interested in studying the actual galaxy itself, so we model it as a small sphere that behaves as a sink and a source. The galaxy has a radius $\rgal~=~0.025\rvir~=~8.0$~kpc~$(\Mh/10^{12} \Msun)^{1/3}$ and defines an effective inner boundary to the active computational domain. When there is an inward mass flux $\dot{\rm M}_{\rm in}$ into the galaxy star formation and feedback are triggered. The star formation rate is a fixed fraction of the inflow rate given by $\dot{\rm M}_\star = \dot{{\rm M}}_{\rm in} ({\eta+1})^{-1}$. `Star formation' in our simulations proceeds by recording and removing $\Delta M_\star = \Delta t  \dot{\rm M}_\star $ from the `galaxy' every time step. The remainder of the newly accreted gas is also removed from the galaxy\footnote{This scheme keeps the gas mass within the galaxy constant, but we allow the thermal pressure of this gas to adjust to match the surroundings thereby avoiding unphysical reflections off of this boundary.}, and is ejected in the galactic wind at a rate given by equation (\ref{eq:eta}). The outflow is launched isotropically\footnote{On time scales of a few Gyr, we find little difference if the gas is ejected isotropically or is confined to fixed opening angles $\sim 60^\circ$; however over the course of $\sim5-10$ Gyr if the direction of the conical outflows is kept fixed the outflows excavate a cavity and blow out along the axis. This may not be realistic because the orientation of a galaxy's outflow will change as its dark matter halo's angular momentum changes over the course of a Hubble time \citep{Book+11,BettFrenk}. Although we adopt an isotropic outflow, instantaneously the outflow often resembles the familiar biconical form as it follows the path of least resistance.} with a velocity $\vw$ that is proportional to the local escape speed\footnote{Our fiducial feedback model has no thermal energy input. When included, thermal energy makes little difference. For the radii at which we inject energy neglecting thermal energy is likely a valid approximation since any hot outflow will have swept up and incorporated a substantial amount of cold gas and adiabatically cooled as it expanded \citep{Thompson+16}.} $\ve$ by adding mass and momentum to cells in a thin shell 1-2 cells wide immediately beyond $\rgal$.
The energetic efficiency of feedback can be expressed with our feedback parameters, $\eta$ and $\vw$, as follows:
\be
\begin{split}
\epsilon_\star &= \frac{\dot{\mathrm{E}}_{\rm out} }{  \dot{\mathrm{M}}_{\star} c^2} = \frac{\frac{1}{2} \dot{\mathrm{M}}_{\rm out} \vw^2 }{ \dot{\mathrm{M}}_{\star} c^2} = \frac{\eta}{2} \left(\frac{\vw}{c}\right)^2\\
 &= 2.7 \times 10^{-6} \left(\frac{\eta}{3}\right) \left( \frac{\vw}{\ve} \right)^2 \left( \frac{\ve}{\rm 400 \,  km\, s^{-1}} \right)^2 .
\end{split}
\label{eq:ep}
\ee

The feedback model parameters are listed in Table \ref{Feedback parameters}. 
For each halo mass we ran a simulation with a high and low mass loading factor. The mass loading factors, $\eta$, were chosen to bracket the expected range for star formation feedback as suggested by observations \citep{Martin99, Veilleux+2005, Heckman+15} and cosmological simulations \citep{Muratov+15}. Likewise the wind velocities span the range of expected velocities of ${\sim}100$s $-10^3$ km s$^{-1}$. In Table~\ref{Feedback parameters} each $\vw$ is listed relative to the escape velocity (of the dark matter halo) at $\rgal$ -- where the wind is launched -- that is defined as
\be
\ve = \sqrt{-2 \Phi_{\rm NFW}(\rgal)} \approx 3.5  v_{\rm vir}.
\label{eq:ve}
\ee
For each choice of $\eta$ in the $10^{11}~\Msun$ haloes we adopted a model with a fiducial $\vw$ and one with a higher $\vw$; we refer to the latter as the `strong' $\vw$ models. These additional models allow us to better study the response of the CGM to the choice of wind model at this halo mass. As we discuss below, the CGM properties are particularly sensitive to changes in feedback in lower mass haloes $\sim 10^{11}~\Msun$.

For reference, we can approximate a standard feedback efficiency by assuming that there is one supernova for every 100 $\Msun$ of stars formed and each supernova supplies $10^{51}$ ergs of energy. The corresponding feedback efficiency is $\epsilon_{\star {\rm, ref}} = 10^{51} \mathrm{ergs} / 100 \mathrm{M}_\odot c^2 = 5.6\times10^{-6}$. Table \ref{Feedback parameters} shows that all of our feedback models have $\epsilon_\star \leq 5.6\times10^{-6}$. Note that we adopt more efficient feedback models in more massive haloes. This is necessary for feedback to have a non-negligible impact on the CGM in these haloes.

\begin{figure}
\hspace*{-0.75cm}   
\includegraphics[width=0.55\textwidth]{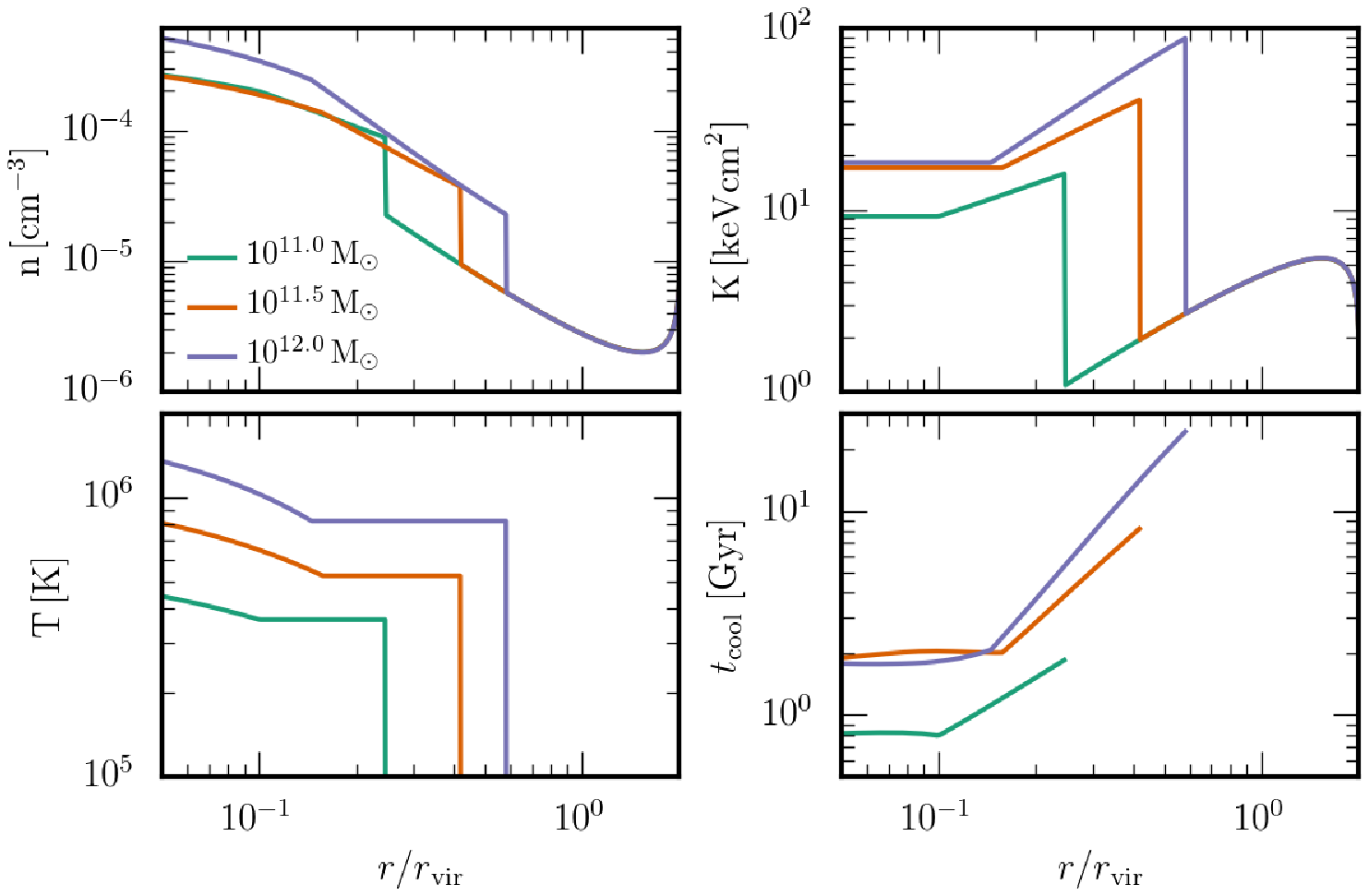}
\caption{ Initial conditions (density $n$, entropy $K=k_B T n^{-2/3}$, temperature $T$, and cooling time $t_{\rm cool}$) for $\Mh = 10^{11},~10^{11.5},$ and$~10^{12}~\Msun$ haloes in teal, orange, and purple, respectively. Initial radii for the isentropic cores are $\rcore/\rvir$ = 0.1, 0.16, and 0.15, and the initial shock radii are $\rsh/\rvir$ = 0.25, 0.42, and 0.58, respectively. Note that the cooling time is not well defined for regions with $T<10^4$ K because we impose a floor to our cooling function at $10^4$ K.}\label{fig:ICs}
\end{figure}

The fiducial initial conditions of the three halo masses we consider, $\Mh = 10^{11},10^{11.5},$ and $10^{12}~\Msun$, are shown in Fig. \ref{fig:ICs}. 
Gas is initialized in a hot virialized halo in hydrostatic equilibrium out to a virial shock radius $\rsh$. 
Beyond the shock and out to $\rta$, the gas is cold ($T_\mathrm{\textsc{igm}}$ = $10^{4}$ K -- the exact temperature of the intergalactic gas does not change the outcome as long as the virial shock is strong, i.e., so long as $T_\mathrm{\textsc{igm}} \ll T_{\rm vir}$) and freely falling with its density set to preserve $\Mdotta$. 
The density and temperature of the gas at the virial shock obey the usual shock jump conditions.
Within the shock the gas is isothermal at the shock temperature until the core radius, $\rcore$, where the gas switches to constant entropy. For the larger mass haloes we consider, there are constraints on properties of the core (e.g., theory: \citealt{MallerBullock, Sharma+12b} and observations: \citealt{Fang+13,Voit+15}), so we choose a consistent value for $\rcore$ -- extrapolating for halo masses with no constraint. For increasing halo mass ($10^{11},10^{11.5},$ and $10^{12}~\Msun$) the initial radii for the isentropic cores are $\rcore/\rvir$ = 0.1, 0.16, and 0.15, and the initial shock radii are $\rsh/\rvir$ = 0.25, 0.42, and 0.58. The corresponding initial gas fractions within $\rvir$ are $f_{\rm gas} = 0.013, 0.017,$ and 0.026.
For comparison, the time average baryon\footnote{Here we define the baryon fraction to be gas between $\rgal$ and $\rvir$ and gas excised from the domain at small radii (`stars').} fractions within $\rvir$ are ${\sim}0.05-0.1$, reflecting the new equilibrium reached after several dynamical times ($t_{\rm dyn} = (G \Mh \rvir^{-3})^{-1/2} = (10 H_\circ)^{-1} = 1.4$ Gyr).

We introduce isobaric density perturbations throughout the domain that break the spherical symmetry. 
The amplitude of the perturbations satisfy $\delta \rho / \rho = 0.3$ and have a power spectrum that goes as $k^{-1/2}$ for $1\leq \frac{k L_{\rm box}}{2 \pi} \leq 100$, where $k$ is the wave number of the perturbation and $L_{\rm box}$ is the size of our domain. 
The results are insensitive to the details of how the perturbations are introduced. We do not add any angular momentum to gas in our domain. We assume that the disk circularizes on small scales ($\lesssim 0.05 \rvir$) \citep[e.g.,][]{Mo+98} comparable to where gas is removed and injected, so angular momentum is not essential on the scales we study here.

To ensure that our results do not depend sensitively on our initial conditions we ran simulations with no initial shock. In this case, a thermal pressure supported gaseous halo never develops in the lower mass $10^{11}~\Msun$ halo because at the `galaxy' radius the accretion shock's cooling time is shorter than all other relevant time-scales.\footnote{The perturbations we impose break spherical symmetry and ensure that there is a shock at small radii.} Alternatively, in higher mass haloes  ($\gtrsim10^{11.5}~\Msun$) an accretion shock at the `galaxy' radius has a sufficiently long cooling time to allow a virialized halo to develop. For all halo masses within a few dynamical times the behavior of simulations with and without initial shocks are very similar.

\begin{figure}
\hspace*{-0.75cm}   
\includegraphics[width=0.525 \textwidth]{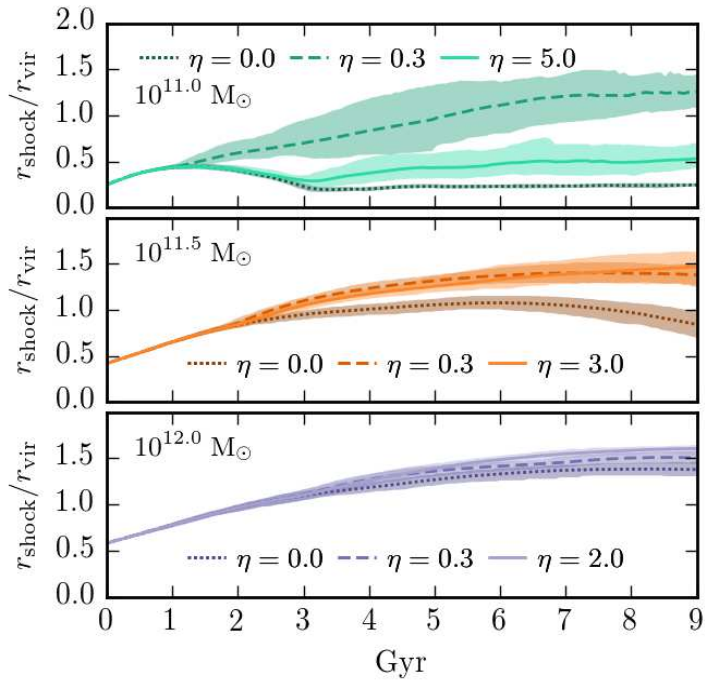}
\caption{The shock radius evolution normalized by the virial radius for simulations with the feedback models listed in Table \ref{Feedback parameters} ($\eta>0$) and with no feedback ($\eta=0$). The `strong' feedback models for the $10^{11}~\Msun$ halo are omitted for clarity. In both of these simulations the shock radius steadily increases reaching 2$\rvir$ by 6 Gyr. The shaded regions show the 1 $\sigma$ quantiles of the shock radii measured at different angles. The initial virial shocks in the higher mass haloes, $\geq 10^{11.5}~\Msun$, gradually grow over time. Alternatively, in the $10^{11}~\Msun$ haloes the initial virial shocks quickly become unstable to cooling and collapse, after which, in the simulations with feedback, incoming gas shocks directly on the outgoing galactic wind -- this is highly aspherical and leads to a range of shock radii for the remainder of the simulation. The shock in the $10^{11}~\Msun$ haloes is best interpreted as a `wind shock' produced when inflowing gas meets outflowing galactic wind material. By contrast at higher masses, the shock is a canonical virial shock between inflow and a roughly hydrostatic halo. The impact of feedback on the longevity of the virial shock decreases with halo mass.}\label{fig:Rshock_Comparison}
\end{figure}

In our simulations we use static mesh refinement in the center of the domain to achieve high spatial resolution in the halo cores. 
The base level resolution is 57 cells per $\rvir$, and our fiducial resolution runs have two additional refined levels, which brings the spatial resolution to $\Delta x = \rvir / 228 = 1.4$~kpc ~$(\Mh/10^{12} \Msun)^{1/3}$. In Appendix \ref{Appendix} we discuss our convergence study. 
We ran these simulations using as many as four levels of refinement and found our primary results well converged. Note, however, that we find a resolution dependence to the inherently non-linear process of cold clump condensation via thermal instability in more massive haloes. This is because the size of the fragments should be $\sim c_s \tcool \approx 0.1$ pc $n^{-1}$ in order for the cooling clouds to remain in pressure equilibrium as they cool \citep{McCourt+16}. Alternatively, if thermal conduction (which we do not include) is not suppressed then the cold clumps should be approximately the Field length $\lambda_F$ (the maximum length scale over which conduction dominates cooling), which is $\lambda_F\lesssim 10 ~\mathrm{pc}$ for $10^{-4}$~cm$^{-3}$ gas at $10^4$ K. In either case the clump size is $\ll \Delta x$, so the cold clump properties are not expected to converge in our simulations.
This resolution dependence may have important implications for cosmological simulations, which may have difficulty resolving the thermal instability in halo gas.

\section{Results} \label{results}
We now present the results of our $10^{11}$ to $10^{12}~\Msun$ halo simulations that use both the high $\eta$ and the low $\eta$ feedback models as listed in Table \ref{Feedback parameters}. 
Simulations of $10^{13}~\Msun$ haloes were similar to $10^{12}~\Msun$ except for requiring a factor of $\sim$2-3 more efficient feedback to suppress runaway cooling.
We focus our attention on how the different feedback models change (or do not change) the resulting CGM structure and the central galaxies' growth. We find that in haloes with $\Mh \gtrsim 10^{11.5}~\Msun$ the CGM properties are relatively insensitive to the choice of feedback model for a wide range of feedback parameters. However, in lower mass haloes, $\lesssim 10^{11.5}~\Msun$, the properties of halo gas depend more sensitively on differences in galactic wind properties.

\begin{figure*}
\hspace*{-0.75cm}   
\includegraphics[width=1.075\textwidth]{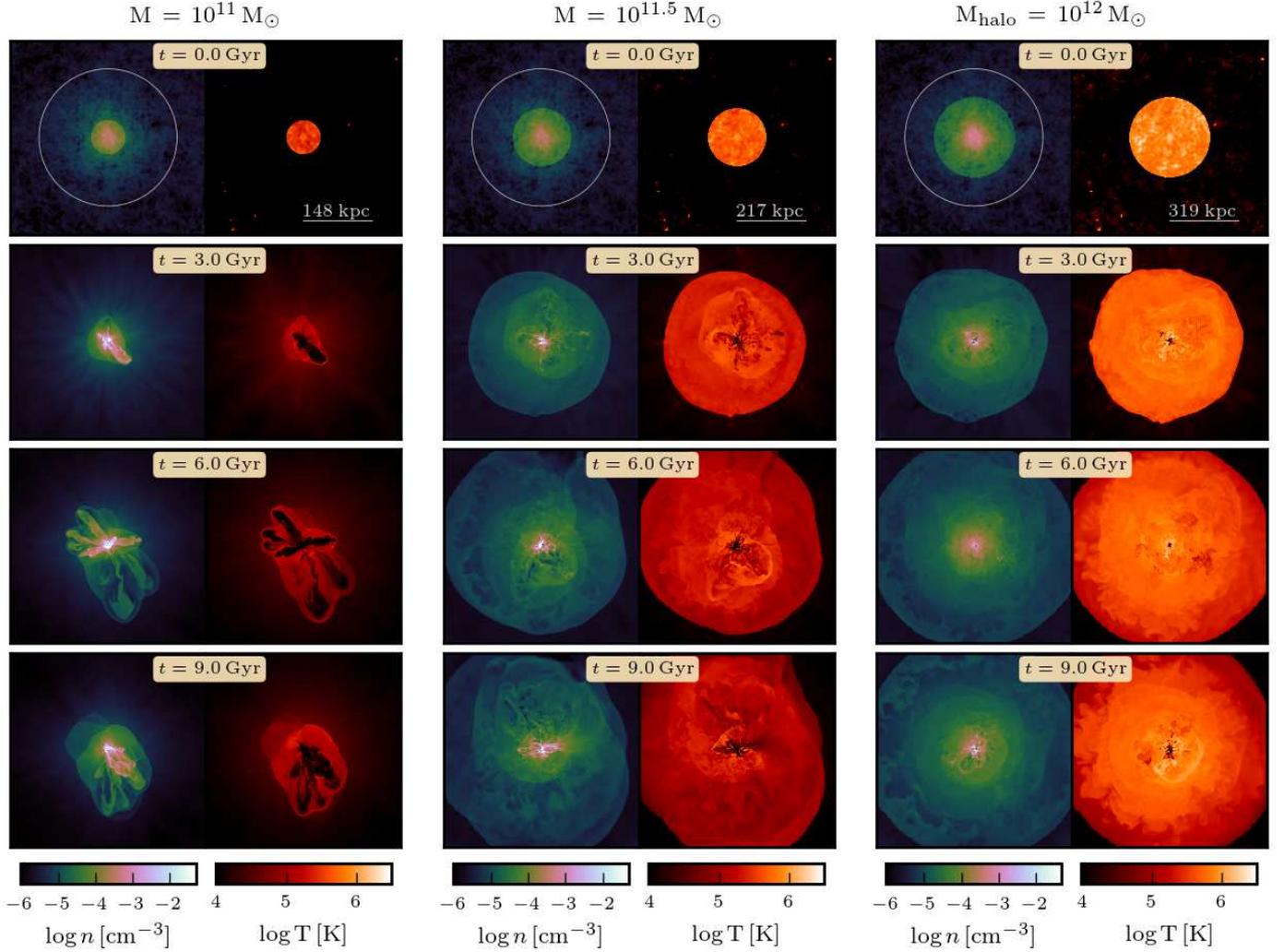}
\caption{ Number density and temperature slices at four times through the center of $10^{11}$, $10^{11.5}$, and $10^{12}~\Msun$ haloes for the fiducial $\eta = 5$, $3$, and $2$ simulations, respectively (Table \ref{Feedback parameters}). The width of each image is 2.8 $\rvir$. The circles and lines in the upper panels have a radius and length of $\rvir$, respectively. In the $10^{11}~\Msun$ halo the initial virial shock quickly collapses and the halo gas transitions to a less spherically symmetric configuration, supported by turbulent motions and ram pressure driven by stellar feedback. In the $10^{12}~\Msun$ halo, the virial shock grows and feedback only affects gas in the core. The effects of changes to the feedback model can be seen in Fig. \ref{fig:M11_slices}, which shows the $10^{11}~\Msun$ halo with a low $\eta$, higher $\vw$ feedback model.}\label{fig:SlicePlots}
\end{figure*}

The cooling rate of astrophysical plasmas ensures that (absent feedback) there is a critical halo mass, $\sim 10^{11.5} - 10^{12} \Msun$, which delineates different physical regimes of circumgalactic gas. Above this critical halo mass gaseous haloes can be thermally supported, but at lower halo masses they cannot \citep{Silk77,ReesOstriker+77,Binney77, BirnboimDekel03}. This is due primarily to the fact that the cooling rate peaks around $10^{5.5}$ K, which is the virial temperature $T_{\rm vir}$ of a halo at $\sim 10^{11.5} - 10^{12}~\Msun$, so the cooling time of the virialized gas is shorter relative to its free-fall time than it is in more massive haloes. Here we show that in the presence of galactic feedback the picture remains similar, but with the modification that in low mass haloes gas can instead be supported by the ram pressure and turbulence generated from vigorous feedback rather than by thermal pressure. The impact of feedback on either side of the critical halo mass is reflected in the evolution, galactic accretion history, and phase structure of the CGM. 

\begin{figure}
\vspace*{-0.5cm}   
\hspace*{-0.5cm}   
\includegraphics[width=0.5\textwidth]{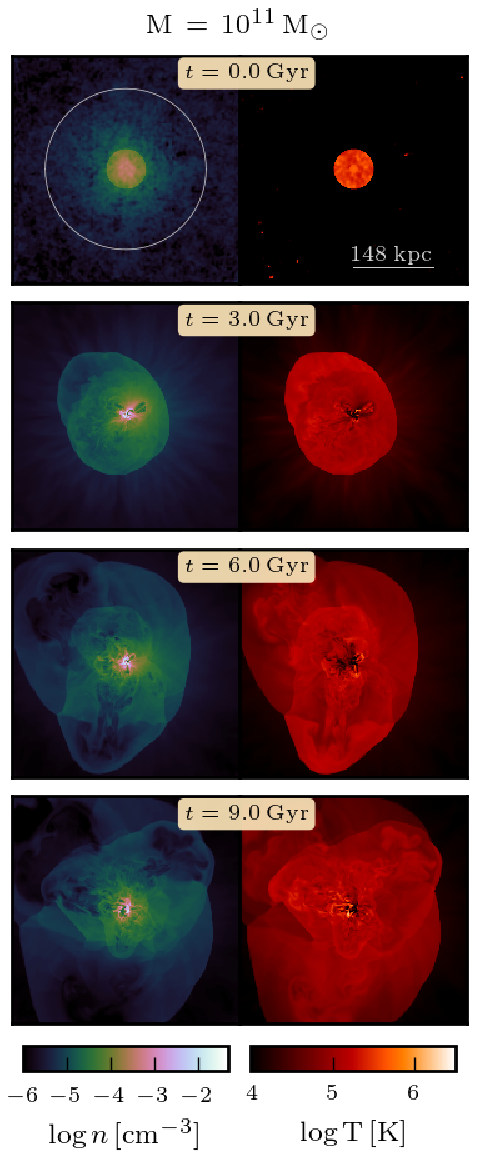}
\vspace*{-0.99cm}   
\caption{ Number density and temperature slices at four times through the center of the $10^{11}~\Msun$ halo with feedback having a higher energy per unit mass (the $\eta=0.3$ and $\vw = \sqrt{4.5} \ve$ model in Table \ref{Feedback parameters}). The width of each image is 2.8 $\rvir$. The circles and lines in the upper panels have a radius and length of $\rvir$, respectively. The lower mass loading and higher wind velocity of this feedback model relative to the high $\eta$ model yield a much hotter and less dense gaseous halo (compare with Fig. \ref{fig:SlicePlots}). We omit the analogous plots for the $10^{11.5}$ and $10^{12}~\Msun$ haloes because in these cases there is visually little difference between the low $\eta$ feedback model and the high $\eta$ feedback model shown in Fig. \ref{fig:SlicePlots}.} \label{fig:M11_slices}
\end{figure}

Fig. \ref{fig:Rshock_Comparison} shows the accretion shock radius\footnote{The accretion shock is identified by a discontinuous drop in the inflow velocity along with an increase in entropy.} evolution in the simulations with both feedback models at all halo masses. To account for the lack of spherical symmetry, we measure the shock radius along 48 equally spaced rays emanating from the galaxy and plot the mean and the 1$\sigma$ range. For comparison we also show the shock radius evolution in simulations without any feedback, i.e., $\eta=0$. Without feedback the change in the CGM properties at the critical halo mass of $\sim 10^{11.5}~\Msun$ is reflected clearly in the differences in shock radii evolution \citep[see also][]{BirnboimDekel03}. However, with feedback this diagnostic is less useful even though the CGM is qualitatively different on either side of the critical halo mass.

In the more massive haloes $\gtrsim 10^{11.5}~\Msun$ the virialized gas remains stable for the duration of the simulations and the shock steadily grows with time. Feedback has a minor impact on the growth of the virial halo in the $10^{12}~\Msun$ halos. 

Going to lower masses the impact of feedback increases. Without feedback the virial shock of the $10^{11.5}~\Msun$ halo begins to collapse after $\sim 6$ Gyr\footnote{In reality, over 6 Gyr this halo may have grown considerably, leading to a deeper potential, higher shock temperature, and less prominent cooling, so this turnover may be an artifact of our non-evolving dark matter potential.}, but with feedback the virial shock radius is relatively insensitive to the choice of feedback parameters. 
In the $10^{11}~\Msun$ haloes the initial virial shock quickly collapses, which is the same with or without feedback. The fact that after the initial collapse there is any shocked gas beyond $\rgal$ is because feedback is driving gas out into the halo, which halts the progression of the inflowing gas. The wider range of shock radii at a given time relative to more massive systems is indicative of the transition from a canonical accretion shock, where inflowing gas hits a roughly hydrostatic, spherical atmosphere, to a `wind' shock, where the inflowing material directly impacts outflowing wind ejecta. At this low halo mass the choice of feedback parameters makes a large difference for the resulting shock evolution. The low $\eta$ model, which has larger wind velocities and wind shock temperatures, is much more effective in halting the advance of large scale accreting gas. Moreover, in both of the `strong' feedback models (with yet larger $\vw$, not shown in Fig. \ref{fig:Rshock_Comparison}), feedback is so efficient that the shock radius expands in certain directions to the outer boundary of the domain thereby entirely halting the inflow of gas and even launching gas beyond $\rta$.

Fig. \ref{fig:SlicePlots} shows density and temperature slices at several different times for haloes with masses of $10^{11}$, $10^{11.5}$, and $10^{12}~\Msun$ with the (fiducial) high $\eta$ feedback efficiency model. In these images the rapid cooling of the initial virialized gas in the $10^{11}~\Msun$ halo is readily apparent. After the virial shock collapse, inflowing gas at this halo mass directly interacts with gas expelled by the galactic wind producing a wind shock. The resulting wind shocks cool quickly because the shock temperature $T_{\rm shock} \sim T_{\rm vir}$ (since $\vw \sim \ve(\rgal) \sim v_{\rm vir}$) and the cooling rate is very high at the virial temperatures for these halo masses -- which is why the initial pressure supported gas collapsed quickly in the first place. These rapidly cooling shocks result in highly anisotropic outflows even though the galactic wind is ejected isotropically. This anisotropy is reflected in the large spread at late times in the measured shock radius in Fig. \ref{fig:Rshock_Comparison}. The outflows change direction on Gyr time-scales and inflowing gas that does not make it all the way to the galaxy is delayed for at most a few dynamical times.

Fig. \ref{fig:SlicePlots} demonstrates that the behavior is strikingly different in the only slightly more massive haloes. The gas in these higher mass haloes ($10^{11.5}$ and $10^{12}~\Msun$) never experiences the dramatic total loss of thermal pressure support. In this case, inflowing gas from large radii is incorporated into the virialized halo via a virial shock and remains far from the central galaxy for many Gyr.  However, cooling does occur in these haloes, but it is primarily in their cores. This cooling leads to inflow and subsequent feedback that in turn stabilizes the halo core against additional cooling.

As demonstrated in the shock radius evolution (Fig. \ref{fig:Rshock_Comparison}), the difference between feedback models is negligible in the higher mass haloes, but in the $10^{11}~\Msun$ haloes the difference can be large. In Fig. \ref{fig:M11_slices} we show density and temperature slices at several different times of the $10^{11}~\Msun$ halo with the fiducial low $\eta$ feedback model ($\eta = 0.3$ and $\vw = \sqrt{4.5} \ve$; see Table \ref{Feedback parameters}). Relative to the fiducial high $\eta$ model, the halo is filled with much more hot, $>10^{5}$~K, and diffuse, $<10^{-4}~\mathrm{cm}^{-3}$, gas. This halo gas remains suspended at large radii, $\sim\rvir$, for much longer than the halo gas in the high $\eta$ model due to its longer cooling time. Visually, the halo gas in the low $\eta$ $10^{11}~\Msun$ simulation in Fig. \ref{fig:M11_slices} resembles the virialized haloes at higher masses in Fig. \ref{fig:SlicePlots}. However, as shown below in Fig. \ref{fig:support}, the halo gas in the low $\eta$ $10^{11}~\Msun$ simulation in Fig. \ref{fig:M11_slices} is not a standard thermally supported halo, but is instead supported to a large extent by bulk motions driven by the galactic wind.

\begin{figure}
\hspace*{-.66cm}   
\includegraphics[width=0.535\textwidth]{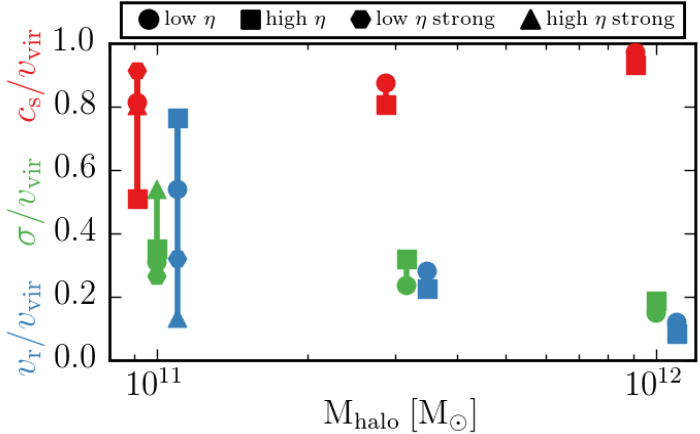}
\vspace*{-0.25cm}   
\caption{Time and mass-weighted radial average of the sound speed (red; shifted to the left by 0.1 dex), velocity dispersion (green), and radial velocity (blue, positive indicates inward motion; shifted to the right by 0.1 dex) normalized by the halo's virial velocity (Table \ref{Feedback parameters}). 
The averaging is done from 3 to 9 Gyrs. Large $c_s/v_{\rm vir}$ indicates the haloes gas is thermally supported, large and positive $v_r/v_{\rm vir}$ indicates the halo gas is primarily freely falling in and under pressurized relative to hydrostatic equilibrium, and large $\sigma/v_{\rm vir}$ indicates turbulent support. The $10^{12}~\Msun$ halo gas is primarily thermal pressure supported, while in lower mass haloes bulk flows (turbulence, inflows, and outflows) are increasingly important.} \label{fig:support}
\vspace*{-0.25cm}   
\end{figure}

Fig. \ref{fig:support} shows the different contributions to the total pressure support in our simulated haloes and encapsulates the primary impact of stellar feedback on the bulk properties of the CGM as a function of halo mass. 
This figure demonstrates that the presence of feedback does not significantly alter the critical halo mass above which thermal pressure supported gaseous haloes can be sustained.  In addition, it shows that below this critical halos mass the halo gas is farther from hydrostatic equilibrium and is supported more by turbulence and bulk flows.   Specifically, Figure 5 shows the density-weighted radial (from $\rgal$ to $\rvir$) and time (from 3 to 9 Gyr) averaged sound speed $c_s$, radial velocity $v_r$ (here we adopt $v_r>0$ for gas flowing toward the center), and velocity dispersion $\sigma=\sqrt{\langle v(r) \rangle^2 - \langle v(r)^2 \rangle}$ (all normalized by the halo virial velocity $v_{\rm vir} = \sqrt{G\Mh \rvir^{-1}}$. To convert to a velocity in km s$^{-1}$, the halo virial velocities $v_{\rm vir}$ are given in Table \ref{Feedback parameters}). 
The sound speed traces the thermal pressure, the velocity dispersion traces turbulent support, and the radial velocity gives a measure of how under ($v_r/v_{\rm vir} > 0$) or over ($v_r/v_{\rm vir} < 0$) pressurized the halo gas is relative to hydrostatic equilibrium.
The strong halo mass dependence of the halo gas dynamics is readily apparent in Fig. \ref{fig:support}.
In the $10^{12}~\Msun$ haloes changes to feedback make little difference to the pressure support. The gas in these haloes is almost entirely thermal pressure supported ($c_s\sim v_{\rm vir}$) and close to hydrostatic equilibrium ($v_r/v_{\rm vir}\sim0$). 
Going to lower masses, the halo gas is farther from hydrostatic equilibrium, the contribution from thermal pressure support decreases, and the contribution of turbulent pressure support increases. Moreover, as we show below for other CGM properties, the sensitivity of the different pressure contributions to the feedback model increases as halo mass decreases.

\begin{figure}
\hspace*{-0.25cm}   
\includegraphics[width=0.525\textwidth]{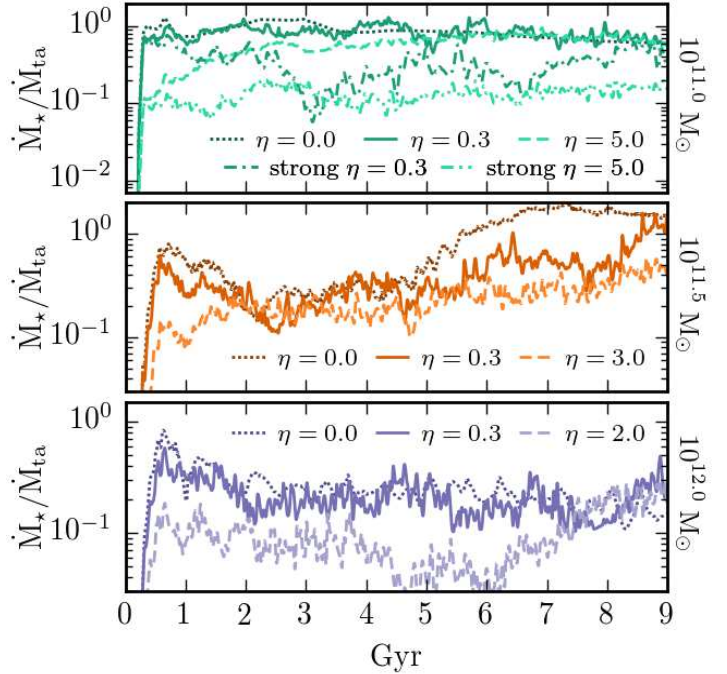}
\caption{ Star formation rate evolution for $10^{11},~10^{11.5},$ and$~10^{12}~\Msun$ haloes, from top to bottom. We show results from no feedback ($\eta=0$), and a range of feedback parameters (see Table \ref{Feedback parameters}). Here the star formation rate is defined to be $\dot{\rm M}_\star = \dot{\rm M}_{\rm in} / (1+\eta)$, where $\dot{\rm M}_{\rm in}$ is the inflow rate at $\rgal$; $\dot{\rm M}_\star$ is also the rate at which gas is removed from the domain at each time step.} \label{fig:Mdots_Plot}
\vspace*{-0.4cm}   
\end{figure}

Fig. \ref{fig:Mdots_Plot} shows the star formation rate normalized by the cosmological accretion rate that is fed into the haloes at the turn around radius, $\Mdotta$, for simulations with and without feedback. 
Recall that we are defining the star formation rate as $\Mdotstar =  \dot{\mathrm{M}}_{\rm in} ({\eta+1})^{-1}$ where $\dot{\mathrm{M}}_{\rm in}$ is the amount of gas that enters the `galaxy,' which we model as a sphere of radius $\rgal = 0.025\rvir$ at the center of the halo.
In the $10^{11}~\Msun$ haloes with the fiducial feedback models even though the galactic winds do arrest some of the inflowing gas, the star formation rate of the galaxy is approximately the same whether or not there is feedback and reaches $\Mdotstar \sim \Mdotta$. With the strong feedback models, however, the powerful wind shock cuts the galactic gas supply, dropping $\Mdotstar$ by an order of magnitude. In the higher mass haloes the stable virial shock at large radii prevents most of the inflowing gas from reaching the galaxy. However, cooling in the halo cores leads to appreciable accretion on to the galaxy. In some cases, feedback at these halo masses suppresses the resulting star formation by up to a factor of ${\sim}2-10$ by reheating the cores. 
{ This is inline with what has been seen in some cosmological simulations that varied the feedback efficiency \citep[e.g.,][]{FaucherGiguere+11, Oppenheimer+10, Nelson+15, Rahmati+15}; However, in simulations of isolated MW-like galaxies, \cite{Marasco+15} found that changing the energy input per supernova by as much as a factor of 32 does not significantly change the star formation rate.}

 \begin{figure}
\hspace*{-1cm}   
\includegraphics[width=0.525\textwidth]{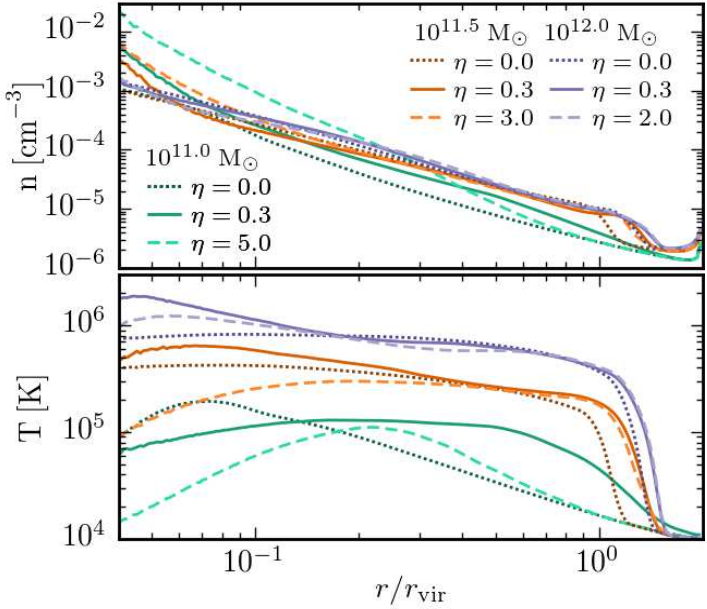}
\caption{ Radial profiles of the number density (top) and temperature (bottom) averaged from 3 to 9 Gyr for all halo masses with and without ($\eta=0$) feedback. The $10^{11}~\Msun$ haloes with the strong feedback models (Table \ref{Feedback parameters}) are omitted for clarity; their profiles are similar to but slightly more diffuse and hotter than the $10^{11}~\Msun$ halo profiles shown. The radial range spans $\sim2\rgal$ to $\rta=2\rvir$. The impact of the feedback model becomes increasingly important in setting the structure of the CGM as halo mass decreases.} \label{fig:Profiles}
\vspace*{-0.25cm}   
\end{figure}

\subsection{Phase structure and dynamics of halo gas}\label{Phase structure and dynamics of halo gas}

Fig. \ref{fig:Profiles} compares the radial profiles of spherically averaged number density and temperature averaged from 3 to 9 Gyr. The $10^{11}~\Msun$ haloes with the strong feedback models are omitted for clarity\footnote{The profiles of the strong feedback $10^{11}~\Msun$ haloes are similar to the fiducial feedback model $10^{11}~\Msun$ haloes, but with slightly lower densities and higher temperatures.}. For all masses the low $\eta$ feedback model (with higher $\vw$) results in lower central densities and higher temperatures, which is in agreement with results from similar numerical experiments \citep[e.g.,][]{Suresh+15a}. The difference between the CGM structure that results from adopting either the low or the high $\eta$ feedback model becomes larger at lower halo masses. The relative insensitivity of the density and temperature profiles to the choice of feedback in the more massive halos agrees with what has been found in cosmological simulations \citep[e.g.,][]{vandeVoort+12}. Additionally, lower mass haloes are more centrally concentrated. The $10^{11}~\Msun$ halo has a density profile power law index ${\lesssim}-2$ -- similar to or steeper than the underlying NFW profile -- and the $10^{12}~\Msun$ halo has a density profile power law index ${\sim}-1.5$. The density profiles of the $10^{12}~\Msun$ haloes in Fig. \ref{fig:Profiles} are in good agreement with what is inferred in the Milky Way from O\textsc{vii} and O\textsc{viii} emission \citep{Miller+15}. 

\begin{figure}
\hspace*{-0.2cm}   
\includegraphics[width=0.55\textwidth]{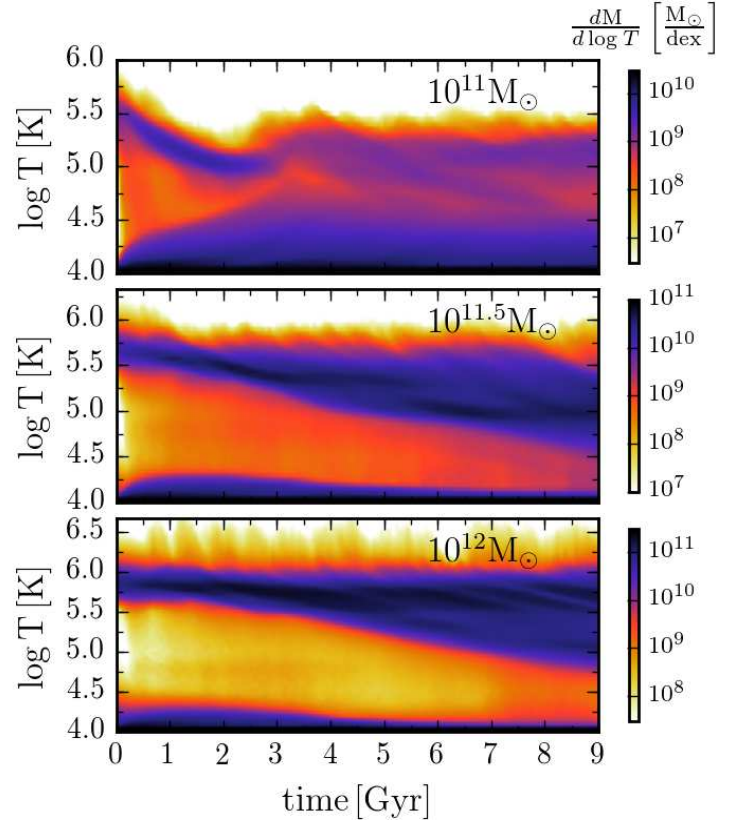}
\vspace*{-0.75cm}   
\caption{The amount of mass per logarithmic temperature bin over time between $2~\rgal$ and $2~\rvir$ is shown for the fiducial high $\eta$ feedback model simulations of all three halo masses. The two higher mass haloes cool slowly leading to modest amounts of $<10^{5.5}$ K gas that cools out of a hotter ambient background. The $10^{11}~\Msun$ halo, on the other hand, cools quickly and all of the $\gtrsim10^{5}$ K gas is a result of the galactic wind shocking on gas accreted from large scales. Fig. \ref{fig:Phase_Evolution_M11} shows that in $\sim 10^{11}~\Msun$ haloes feedback with a larger energy per unit mass (lower $\eta$; Table \ref{Feedback parameters}) leads to a much broader phase distribution in the CGM.}\label{fig:Phase_Evolution}
\end{figure}

None of our haloes demonstrate a clear density core. Such cores are often used in phenomenological modeling of halo gas \citep[e.g.,][]{MallerBullock,Sharma+12b,Voit+15b}. Previous studies with a similar approach to ours that focused on slightly more massive haloes ($\gtrsim 10^{13.5}~\Msun$) found distinct density cores in their haloes at a radius of $\lesssim 0.05 \rvir$ \citep{Sharma+12a}. The lack of cores in our haloes may be a consequence of insufficient resolution close to the inner edge of our domain ($\rgal = 0.025 \rvir$), or the limited region of the $\eta-\vw$ parameter space covered by our models. In particular, the feedback in our simulations, which is in the form of a galactic wind, tends to produce a roughly $r^{-2}$ density profile due to either inflow or outflow at small radii. Yet lower $\eta$ and higher $\vw$, or thermal feedback -- as was used by \cite{Sharma+12a} -- may be required to produce a significant density core.

\begin{figure}
\hspace*{-0.2cm}   
\includegraphics[width=0.55\textwidth]{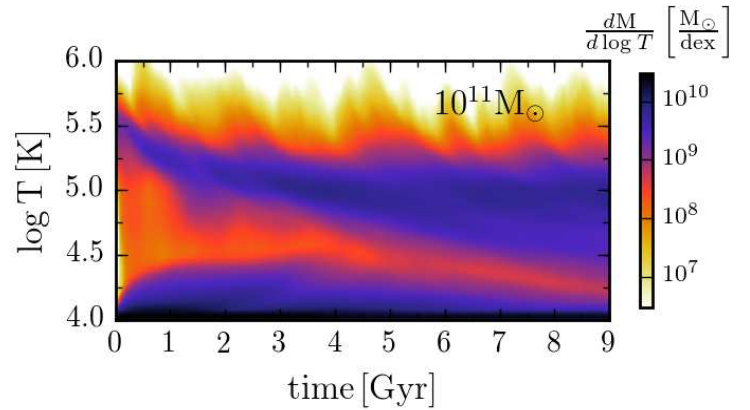}
\vspace*{-0.75cm}   
\caption{ Amount of mass per logarithmic bin in $T$ for the $10^{11}~\Msun$ halo with the fiducial low $\eta$ feedback model (Table \ref{Feedback parameters}). The lower mass loading and higher wind velocity of this feedback model lead to a substantial amount of warm gas ($10^5-10^6$ K); compare with Fig. \ref{fig:Phase_Evolution}. The gas in this temperature regime is a result of galactic wind shocks and the resulting rapid cooling, not an accretion shock on to a static halo as is the case in $\gtrsim 10^{11.5}~\Msun$ haloes.} \label{fig:Phase_Evolution_M11}
\vspace*{-0.25cm}   
\end{figure}

\begin{figure}
\hspace{-0.5cm}   
\includegraphics[width=0.525\textwidth]{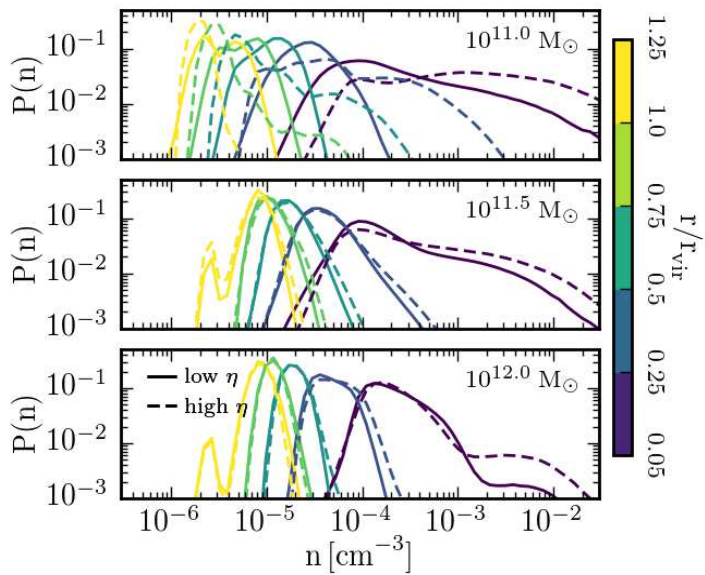}
\caption{ The time averaged (from 3 to 9 Gyr), mass weighted density probability distribution function in radial bins. The yellow line corresponds to 1.25 to 1 $\rvir$, the green line to 1 to 0.75 $\rvir$, the teal line to 0.75 to 0.5 $\rvir$, the blue line to 0.5 to 0.25 $\rvir$, and the purple line to 0.25 to 0.05 $\rvir$ = 2 $\rgal$. Halo mass increases from top to bottom, and the low and high $\eta$ feedback models are plotted with solid and dashed lines, respectively. The $10^{11}~\Msun$ haloes with the strong feedback models are omitted because they are similar to the fiducial low $\eta$ model shown here. The peak around $2\times10^{-6}$ cm$^{-3}$ corresponds to undisturbed cosmologically accreted gas. The $10^{11}~\Msun$ halo gas has a broader range of densities at a given radius, due to the larger impact of stellar feedback on the CGM.} \label{fig:PDF}
\end{figure}

Figures \ref{fig:Phase_Evolution} and \ref{fig:Phase_Evolution_M11} show a different way of quantifying the thermal structure of the halo gas, via $d \mathrm{M}/d \log\mathrm{T}$, the mass contained in a given logarithmic bin in temperature. The radial range extends from 2$\rgal$ out to the outer boundary of the domain 2$\rvir$. Quantifying the amount of mass in different temperature regimes is directly related to many of the best observational constraints we have on the structure of the CGM. X-ray emission and absorption are sensitive to gas at $\geq 10^6$ K (e.g., Milky Way: \citealt{Gupta+12,Miller+16}, other galaxies: \citealt{Forman+85,OSullivan+01,Mulchaey+10,AndersonBregman+11}).
In addition, UV absorption lines in the spectra of background quasars in dark matter haloes at these masses can be used to measure the amount of mass in different temperature regimes $\sim 10^4 - 10^{5.5}$ K \citep[e.g.,][]{Steidel+10, Tumlinson+11, Rudie+12, Werk+14, Borthakur+15}.
Fig. \ref{fig:Phase_Evolution} shows the phase structure evolution for the (fiducial) high $\eta$ feedback models and for comparison Fig. \ref{fig:Phase_Evolution_M11} shows the same quantity for the $10^{11}~\Msun$ halo with the fiducial low $\eta$ feedback model (the differences between the feedback models in the two higher mass haloes are minor so they are not shown). The phase structure in the $10^{11}~\Msun$ haloes with the strong feedback models (see Table \ref{Feedback parameters}) are similar to that shown in Fig. \ref{fig:Phase_Evolution_M11}. 

With the fiducial high $\eta$ feedback model the vast majority of the mass resides at T $\lesssim 10^{4.5}$ K in the $10^{11}~\Msun$ halo, whereas with the low $\eta$ feedback model the wind is able to populate the intermediate temperature range, $\sim 10^{4.5}-10^{5.5}$ K, with significant amount of gas. This is because of the longer cooling times of the wind shock heated gas with the larger $\vw$ in the low $\eta$ model. In the $10^{11.5}$ and $10^{12}~\Msun$ haloes the majority of the mass is at the virial temperature $\sim10^{5.5}-10^6$ K. Cooling of the accretion shock heated gas, the formation of dense clumps by thermal instability, and cooling of galactic wind shocks eventually fill the intermediate temperature range.
The origin of the intermediate temperature halo gas thus differs dramatically in haloes above and below $10^{11.5}~\Msun$. Moreover, below $10^{11.5}~\Msun$ the amount of gas in a given temperature regime -- particularly the cool/warm $\sim 10^{4.5}-10^{5.5}$ K regime is more sensitive to the feedback model. Alternatively, in the haloes with long lived thermal pressure support, $\gtrsim 10^{11.5}~\Msun$, the amount of gas in a given temperature regime depends more on the mass of the halo and less (although non-negligibly) on the feedback physics. 

The inhomogeneous density structure of our haloes can be seen in Fig. \ref{fig:PDF} which shows the time averaged (from 3 to 9 Gyr), mass weighted density probability distribution function in radial bins extending from 2$\rgal$ to 1.25$\rvir$. Changing the range of times for the averaging makes essentially no difference, as long as a few dynamical times have elapsed, which allows any initial transients to pass (this is true for all of the time averaged plots we show despite the lack of a true equilibrium in Figures \ref{fig:Phase_Evolution} and \ref{fig:Phase_Evolution_M11}). The $10^{11}~\Msun$ haloes with the strong feedback models are similar to the fiducial low $\eta$ model, so they are omitted. The $10^{11}~\Msun$ haloes have significant amounts of mass in a broad range of densities extending from $\sim10^{-6}$ cm$^{-3}$ -- predominantly at large radii $\sim\rvir$ -- to $\gtrsim10^{-2}$ cm$^{-3}$ -- mostly in the halo cores. In a given radial bin the width of the density distribution spans more than an order of magnitude, and even relatively close to the galaxy, $\sim 0.5 \rvir$, there is an appreciable amount of gas at the low densities ($\sim 10^{-5}$ cm$^{-3}$) where photoionization dominates over collisional ionization. 
Going to higher masses the width of density distribution at a given radius shrinks, as do the differences that result from the different feedback models; this is also reflected in the density profiles shown in Fig. \ref{fig:Profiles}. 

\begin{figure}
\hspace*{-0.5cm}   
\includegraphics[width=0.525\textwidth]{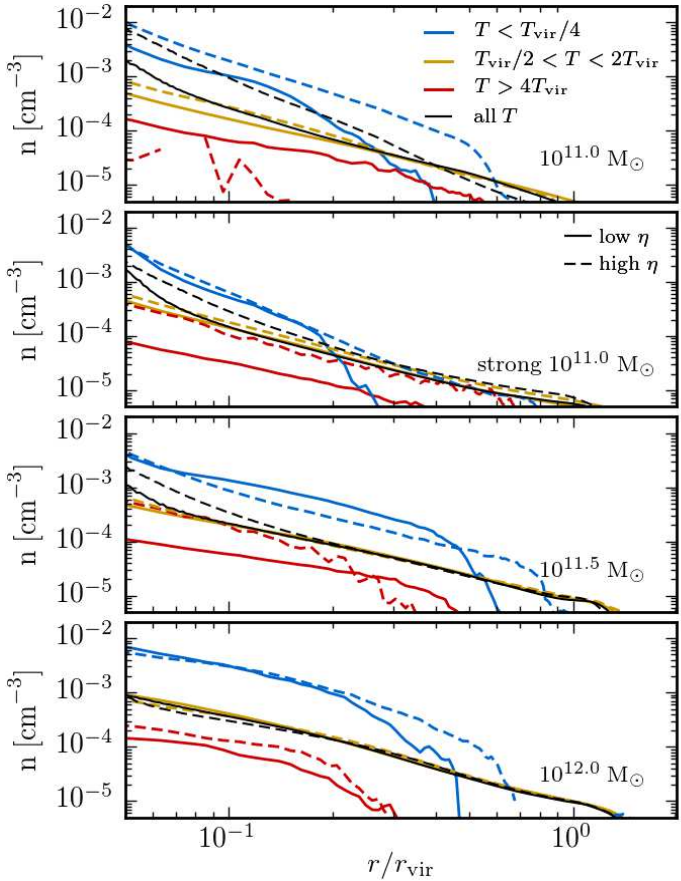}
\caption{The time averaged (from 3 to 9 Gyr) number density profiles of gas in three temperature bins. The hot gas with $T>4T_{\rm vir}$, virialized gas with $T_\mathrm{vir}/2<T<2T_\mathrm{vir}$, and cold gas with $T<T_\mathrm{vir}/4$ are shown in red, gold, and blue, respectively. For reference, the black lines show the number density profiles for gas at all temperatures -- same as in Fig. \ref{fig:Profiles}. The solid (dashed) lines correspond to the low (high) $\eta$ feedback models.} \label{fig:n_profile_Tbins}
\vspace*{-0.25cm}   
\end{figure}

\begin{figure*}
\hspace*{-0.5cm}   
\includegraphics[width=1\textwidth]{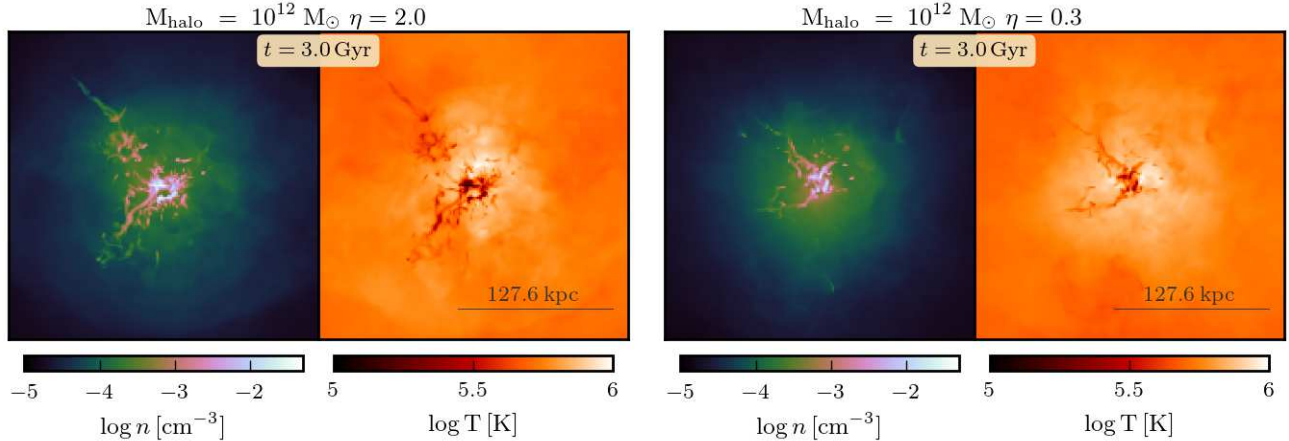}
\vspace*{-0.5cm}   
\caption{ The line-of-sight density weighted average number density and temperature in the centers of the two $10^{12}~\Msun$ haloes after 3 Gyr of evolution. Cold, dense clumps are evident within $\sim100$ kpc of the central galaxy. The left (right) two images are from the high (low) $\eta$ feedback simulations. The images are $0.8\rvir = 255$ kpc across.} \label{fig:M12_projection}
\vspace*{-0.5cm}   
\end{figure*}

\begin{figure}
\hspace*{-0.65cm}   
\includegraphics[width=0.5\textwidth]{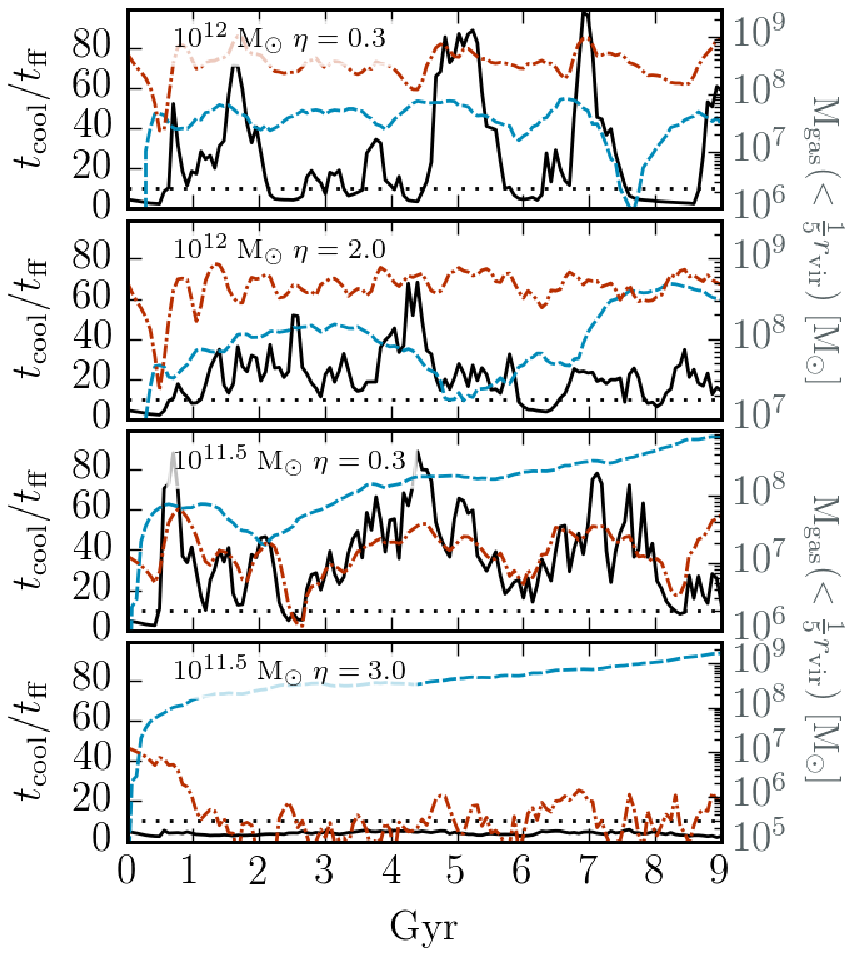}
\caption{Spherical shell averaged $\tcool/\tff$ evolution at $r=3\rgal=0.075~\rvir$ for the $10^{12}~\Msun$ halo simulation with the low $\eta$ (top) and high $\eta$ (upper middle) feedback model, and for the $10^{11.5}~\Msun$ halo simulation with the low $\eta$ (lower middle) and high $\eta$ (bottom) feedback model. We measure $\tcool$ in the hot gas ($>T_{\rm vir}$) only. A thin dotted line is drawn at  $\tcool/\tff=10$, the value below which thermal instability is predicted to lead to multiphase gas and large accretion rates of cold gas.
Additionally, in each panel we plot the amount of cold ($<10^{4.5}$ K, blue dashed line) and hot ($>10^{6}$ K, red dot-dashed line) gas contained within $0.2~\rvir$. In the top three panels the core gas spends most of its time with $\tcool/\tff > 10$ with occasional forays down to $\tcool/\tff \lesssim 10$. 
By contrast, in the $10^{11.5}~\Msun$ halo with high $\eta$, $\tcool/\tff < 10$ at all times. 
In the $10^{12}~\Msun$ halo with low $\eta$ (top) changes to the cold gas mass correlate particularly well with $\tcool/\tff$, a signature of thermal instability.} \label{fig:tcool_tff_evolution}
\end{figure}

Fig. \ref{fig:n_profile_Tbins} shows the time averaged (from 3 to 9 Gyrs) density profiles for gas in three temperature bins, which are delineated relative to the haloes' virial temperatures. We also reproduce the density profiles for all of the gas as shown in Fig. \ref{fig:Profiles}. For reference, the virial temperature is $T_{\rm vir} = 1.1\times10^{5}\mathrm{K},~2.3\times10^{5}\mathrm{K}$, and $5.1\times10^{5}\mathrm{K}$ for the $10^{11}$, $10^{11.5}$, and $10^{12}~\Msun$ haloes, respectively. 
For each halo the super-virial temperature bin is populated by gas that has been shock heated when the wind material interacts with the ambient CGM. The density of this hot gas in $\leq~10^{11.5}~\Msun$ haloes is sensitive to the feedback parameters. The low $\eta$ models have higher wind shock temperatures because of their higher $\vw$ and they also have lower wind densities by definition. This results in more volume filled by the super-virial gas. In addition, in the low $\eta$ models this hot gas extends to larger radii, but has lower density, than in the high $\eta$ models\footnote{Note that this is true in the `strong' $10^{11}~\Msun$ haloes, but at densities below those shown in Fig. \ref{fig:n_profile_Tbins}.}. Observations of gas in this super-virial temperature range would therefore be very useful in constraining the properties of galactic winds; however, to date, most CGM observations are sensitive to gas at $T<10^{5.5}\mathrm{K}\lesssim T_{\rm vir}$.

Fortunately, the cold gas properties also change in key ways with halo mass and feedback model, which should allow existing and future observations of the cold CGM to constrain galactic wind properties. Focusing on the density profiles of the cold gas in Fig. \ref{fig:n_profile_Tbins} (blue line; $T<T_\mathrm{vir}/4$) it is clear that in all cases the high $\eta$ model result in cold gas out to larger radii. We can gain even more insight and constraining power by comparing the cold gas profiles of the more massive haloes ($10^{12}~\Msun$) that have almost entirely virialized hydrostatic haloes to those of the lower mass haloes ($10^{11}~\Msun$) that have more rapid cooling and more vigorous winds and turbulence. In the high mass haloes the cold gas has very high central densities ($\sim 10^{-2}$ cm$^{-3}$) that decreases slowly with radius until it reaches a sharp cut off at a few~$\times0.1\rvir$. By contrast, in the lower mass haloes the cold gas typically has a slightly lower central density ($\sim 3\times10^{-3}$ cm$^{-3}$) that decreases more quickly with radius. In the intermediate halo mass of $10^{11.5}~\Msun$ the cold gas profile with the low $\eta$ feedback model resembles that of the higher mass haloes, while the high $\eta$ feedback model results in a profile that resembles that of the lower mass haloes. Physically, the difference between these profiles is due to the degree of pressure confinement. In the higher mass haloes the cold gas is predominantly surrounded by much hotter, confining gas that drives its density up (see Fig. \ref{fig:M12_projection} for an example density and temperature maps that show these pressure confined cold clumps). This is true for both the cold gas launched by the wind and for cold gas that forms as a result of thermally instability. It is worth noting that this cold gas is under-pressurized relative to the virialized gas by a factor of $\sim2-3$; however, these cold clumps are only marginally resolved in our simulations so we avoid drawing too strong conclusions from this fact (see Fig. \ref{fig:Pressure_Equilibrium_convergence}).

In contrast to the higher mass haloes, in the lower mass haloes the rapid cooling of virialized gas diminishes the thermal pressure of the confining medium while continually driving gas down to low temperatures. This populates the low temperature regime without forcing the densities of the cold gas up. Additionally, the rapid cooling even at large radii and the vigorous feedback triggered by the accretion of cold gas drives cold gas out to large radii. This interpretation of the differences in cold gas profiles is supported by Fig. \ref{fig:support}, which shows that there is more virialized, pressure confining gas in higher mass haloes, and more energy in bulk flows to support cold gas at large radii in lower mass haloes.

\subsection{Thermal instability in the cores of massive haloes}\label{subsection:TI}

In the more massive haloes ($\gtrsim 10^{11.5}~\Msun$) feedback does little to modify the bulk of the CGM out near $\rvir$. However, in their cores the cooling times can be significantly shorter than both a Hubble time and the duration of the simulation. This short cooling time leads to significant inflowing gas and star formation. As is shown Fig. \ref{fig:Mdots_Plot} the resulting galactic wind can in some cases lowers the star formation rate by a factor of a few up to an order of magnitude. An analogous scenario occurs in group and cluster mass haloes ($\Mh \gtrsim 10^{13.5} \Msun$). In these systems central cooling times imply large star formation rates that are inconsistent with observations and point to a heating source that is capable of preventing a cooling flow (e.g., \citealt{McNamaraNulsen2007}). 
At this higher halo mass the central heating source is usually assumed to be an active galactic nucleus (AGN), fueled by gas cooling out of the hot halo. The net inflow of cool gas is significantly larger when the hot halo gas is thermally unstable, which requires $\tcool/\tff \lesssim 10$ \citep{McCourt+12,Sharma+12a, Li+15}. When this condition is satisfied the cold phase rains out on to the central galaxy and triggers enough feedback to reheat the ambient medium and extend the cooling time before the full cooling flow develops. Gas continues to rain out until the feedback drives $\tcool/\tff > 10$. Much of the work to date on this global feedback regulation of hot haloes has focused on more massive systems than we consider here and in the regime of the $\eta$-$\vw$ parameter space appropriate for AGN feedback -- lower $\eta$ and higher $\vw$. This same sort of thermal instability regulation \emph{may} occur in the halo mass range we consider. Indeed, Fig. \ref{fig:M12_projection}, which shows the presence of cold, dense clumps in the cores of the $10^{12}~\Msun$ simulations, seems to demonstrate this thermal instability triggered precipitation.
 
To quantitatively assess the role of thermal instability Fig. \ref{fig:tcool_tff_evolution} shows the time evolution of $\tcool/\tff$ for the hot gas with $T>T_{\rm vir}$ at $3\rgal=0.075\rvir$ in the $10^{11.5}$ and $10^{12}~\Msun$ halo simulations with both feedback models (We omit the $10^{11}~\Msun$ haloes because they are similar to the high $\eta$ $10^{11.5}~\Msun$ halo in that the hot gas cooling time is always less than its free fall time). Also shown is the amount of cold ($<10^{4.5}$ K, blue dashed line) and hot ($>10^{6}$ K, red dot-dashed line) gas contained within $0.2\rvir$. In all but the high $\eta$ feedback $10^{11.5}~\Msun$ halo, $\tcool/\tff >10$ most of the time with occasional dips below $\sim10$. 
Distinguishing cold clump condensation due to thermal instability and cooling triggered by wind shocks (or other sources of cold gas) is non-trivial (which may explain the similarity in Fig. \ref{fig:M12_projection} and the difference in Fig. \ref{fig:tcool_tff_evolution}). 
Note, however, that the cold gas content in the low $\eta$ feedback model $10^{12}~\Msun$ halo only rises when $\tcool/\tff < 10$, and that these increases precede an increase in hot gas. These are strong indicators of the same type of thermal instability regulation as is seen in simulations of high mass haloes. 
In the other simulations that have $\tcool/\tff >10$ (the high $\eta$ $10^{12}~\Msun$ and the low $\eta$ $10^{11.5}~\Msun$ haloes) this correlation is less obvious, so definitively determining if thermal instability triggered feedback plays any role is more difficult. 
The haloes that do not show the clear signs of thermal instability have more vigorous turbulence in their cores (see Fig. \ref{fig:support}). Rapid turbulent mixing relative to the thermal instability growth time-scale can render the instability ineffective \citep{Parrish+10,RuszkowskiOh10}. 
The increased turbulence is tied to the fact that the feedback is primarily kinetic rather than thermal. In the simulations of thermal instability triggered feedback in more massive haloes the feedback models have relatively high energy per unit mass or are purely thermal \citep[e.g.,][]{Sharma+12a,Gaspari+12,Li+15}. The absence of clear signatures of thermal instability triggered feedback in our haloes with more turbulent supported cores may indicate the need for thermal feedback for this regulation to work.

The hot gas in the cores of the $10^{11}~\Msun$ haloes and in the high $\eta$ $10^{11.5}~\Msun$ halo never has $\tcool/\tff \gtrsim 10$ (for much of the time $\tcool\lesssim\tff$), which demonstrates that thermal instability triggered feedback is not a dominant process. Therefore, models of the impact of thermal instability on galaxy formation should not be extended to lower mass haloes, $\lesssim 10^{12}~\Msun$ \citep{Voit+15b}.

\subsection{Connection to quasar absorption observations}\label{Observations}
Here we briefly present additional analysis of our simulations for comparison to quasar absorption observations of the $z\sim0$ CGM \citep[e.g.,][]{Tumlinson+13,Stocke+13}. These studies have enabled measurements of the column density of low-ionization state metals and neutral hydrogen at $T \sim 10^4$ K \citep{Werk+14}, intermediate ionization state gas at $T \sim 10^5$ K \citep{Bordoloi+14}, and higher ionization state gas at temperatures up to $\sim10^{5.5}$ K \citep{Tumlinson+11}. Additional information about the halo gas kinematics can be gleaned from the line widths and velocity offset of the absorbing gas relative to its host galaxy \citep[e.g.,][]{Werk+16}.

\begin{figure}
\hspace*{-0.5cm}   
\includegraphics[width=0.525\textwidth]{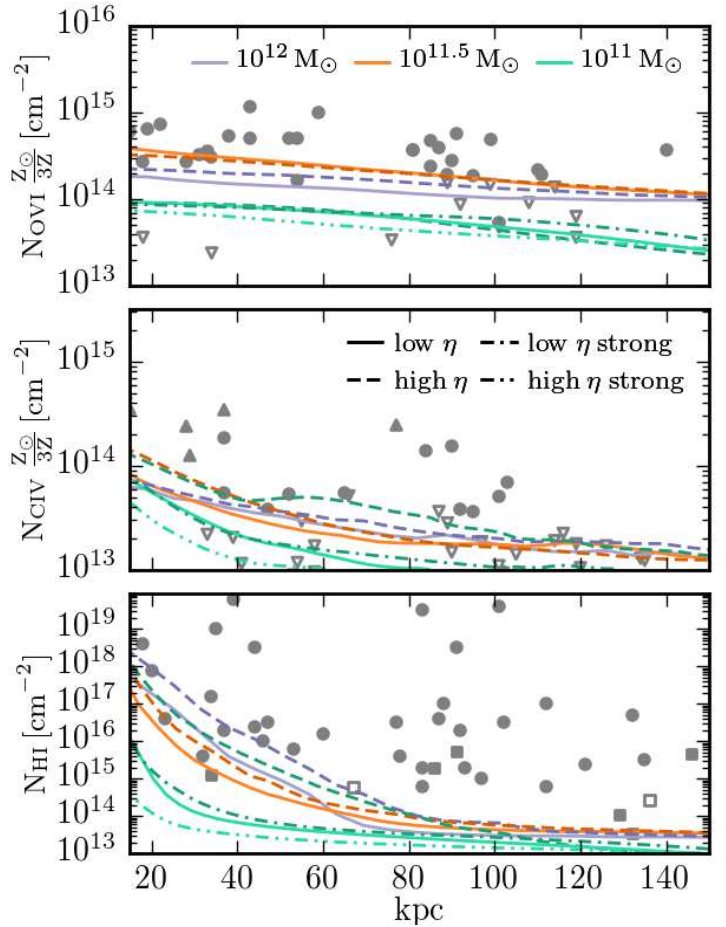}
\vspace*{-0.5cm}   
\caption{ O\thinspace\textsc{vi} (top), C\thinspace\textsc{iv} (middle), and H\thinspace\textsc{i} (bottom) column density profiles in units of cm$^{-2}$ for all of our simulated haloes, time averaged from 3 to 9 Gyrs. Our simulations assume a fixed third solar metallicity everywhere so the inferred O\thinspace\textsc{vi} and  C\thinspace\textsc{iv} column densities can be roughly (not exact because cooling changes) scaled up or down proportional to the metallicity. The points show observations, unfilled symbols represent upper limits, and upward triangles represent lower limits. The O\thinspace\textsc{vi} observations are from Tumlinson et al. (2011), the C\thinspace\textsc{iv} observations are from Bordoloi et al. (2014), and the H\thinspace\textsc{i} come from Werk et al. (2014) (circles) and Prochaska et al. (2011) (squares). }\label{fig:OVI_HI_Column_Prof}
\vspace*{-0.5cm}   
\end{figure}

\begin{figure*}
\hspace*{-0.5cm}   
\includegraphics[width=1.05\textwidth]{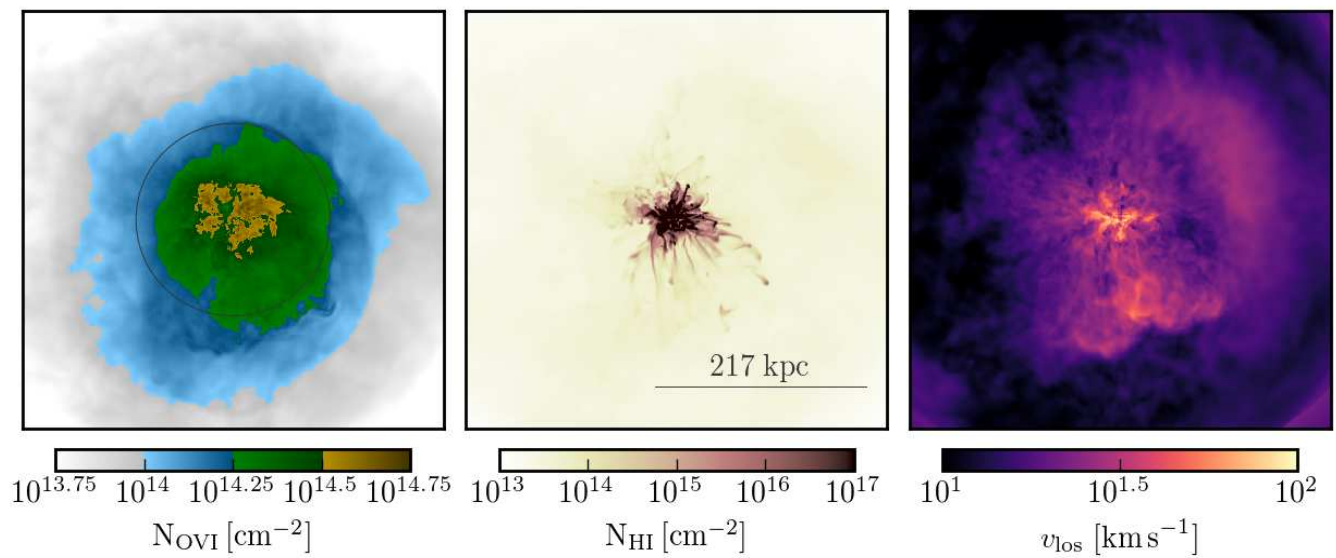}
\vspace*{-0.75cm}   
\caption{Maps of the O\thinspace\textsc{vi} (left) and H\thinspace\textsc{i} (middle) column density, and density weighted average line-of-sight velocity (right) from the $10^{11.5}~\Msun$ halo with the high $\eta$ feedback model after 6 Gyr of evolution. The circle in the left panel has a radius of 100 kpc, and the line in the middle panel is $\rvir$ long. All three images are $2 \rvir$ = 434 kpc across. } \label{fig:OVI_HI_Column}
\vspace*{-0.25cm}   
\end{figure*}

The simplifications inherent to our idealized setup make detailed comparisons between our results and observations suspect. However, a rough comparison can be fruitful, particularly for understanding the trends with changes in halo mass and feedback models. 
The gas fractions of our simulated haloes at late times range from $\sim0.15~(10^{11}~\Msun)$ to $\sim0.25~(10^{12}~\Msun)$ of the cosmic baryon fraction, only a factor of $\lesssim2$ below what is found at $z=0$ in cosmological simulations with stellar feedback \citep[e.g.,][]{vandeVoort+16}. To compare our halo gas properties to observations in more detail, Fig. \ref{fig:OVI_HI_Column_Prof} shows the O\thinspace\textsc{vi}, C\thinspace\textsc{iv} and H\thinspace\textsc{i} column densities averaged from 3 to 9 Gyrs in all of our simulated haloes with feedback. The ionization state of the oxygen and carbon, which depends on temperature and density, is calibrated to ionization equilibrium models calculated using \textsc{cloudy} (\citealt{CLOUDY}). The neutral hydrogen fraction is approximated using an analytic fit to full radiative transfer simulations \citep{Rahmati+13}. Fig. \ref{fig:OVI_HI_Column} shows an O\thinspace\textsc{vi} and an H\thinspace\textsc{i} column density map, as well as a density weighted line-of-sight velocity map for the high $\eta$ $10^{11.5}~\Msun$ halo.
Recall that our simulations have a fixed third solar metallicity throughout the domain. For this reason, we explicitly add the metallicity dependence to the average O\thinspace\textsc{vi} and C\thinspace\textsc{iv} column density profiles shown in Fig. \ref{fig:OVI_HI_Column_Prof}.

The O\thinspace\textsc{vi} column density profiles shown in the top panel of Fig. \ref{fig:OVI_HI_Column_Prof} are similar to those measured by \cite{Tumlinson+11} who found that galaxies presumed to be residing within $\sim 10^{12}~\Msun$ haloes have $N_{\rm O\thinspace\textsc{vi}} > 10^{14.25}$ cm$^{-2}$ out to $\lesssim 150$~kpc. Our simulated haloes in that same mass range have column densities on the lower cusp of the observations differing by a factor of $\lesssim 2$. Additionally, the observed covering fraction of O\thinspace\textsc{vi} gas is ${\sim}0.8-1$, which is in excellent agreement with our $10^{11.5}$ and $10^{12}~\Msun$ haloes. The O\thinspace\textsc{vi} column density map in the left panel of Fig. \ref{fig:OVI_HI_Column} demonstrates this high cover fraction. 
We find that the O\thinspace\textsc{vi} column density peaks in haloes with mass $\sim10^{11.5}~\Msun$. A recent study that accounted for non-equilibrium ionization found that the observed star formation rate dependence of the halo O\thinspace\textsc{vi} content may be primarily driven by a halo mass dependence, similar to what we find, albeit peaking at slightly higher halo masses of ${\sim}10^{12}~\Msun$ \citep{Oppenheimer+16}. Alternatively, \cite{Suresh+15b} found in their simulations that the star formation dependence of the halo O\thinspace\textsc{vi} content was due to the heating and removal of halo gas by AGN feedback.

Fig. \ref{fig:OVI_HI_Column_Prof} shows that the $\sim10^{11.5}~\Msun$ haloes' O\thinspace\textsc{vi} column densities have essentially no dependence on the feedback model. This is because the fraction of collisionally ionized oxygen in the O\thinspace\textsc{vi} state peaks at $T\approx 10^{5.5}$K, which is $\sim T_{\rm vir}$ at this halo mass. Therefore, in $\sim10^{11.5}~\Msun$ haloes O\thinspace\textsc{vi} traces the virialized gas, and, as we have shown, changes to the feedback model have little impact on the amount and structure of virialized gas at these halo masses. 
In the higher mass haloes, on the other hand, O\thinspace\textsc{vi} instead traces gas that is somewhat cooler than the virialized gas, so it is gas that has cooled out of the hotter medium or has been launched out of the galaxy by the wind. O\thinspace\textsc{vi} in $\sim10^{11}~\Msun$ haloes traces gas that is above $T_{\rm vir}$, so it's presence is entirely due to feedback, which explains why the O\thinspace\textsc{vi} column density depends more sensitively on feedback at this halo mass. Interestingly, \cite{Liang+16} found that increasing the strength of their feedback in their cosmological zoom-in simulations of $\sim 10^{12}~\Msun$ halos increased the O\thinspace\textsc{vi} column density dramatically \citep[see also, e.g.,][]{Hummels+13}. This seems to be a result of enhanced metal enrichment with stronger feedback, which would not be captured in our constant metallicity simulations.

The fraction of carbon in the C\thinspace\textsc{iv} ionization state peaks at $\sim10^5$K. The C\thinspace\textsc{iv} column density profiles shown in the middle panel of Fig. \ref{fig:OVI_HI_Column_Prof} are similar to what was measured by \cite{Bordoloi+14} who found that $10^{11}-10^{11.5}~\Msun$ haloes have $N_{\rm C\thinspace\textsc{iv}} > 10^{13.5}$ cm$^{-2}$ out to $\lesssim 100$~kpc. As with the O\thinspace\textsc{vi}, the C\thinspace\textsc{iv} column densities of our haloes are on the low end of the observations, but given the simplifications of our simulations and the assumption of constant metallicity the agreement is encouraging. The most striking feature of the C\thinspace\textsc{iv} column profiles is the large differences between the profiles of the $10^{11}~\Msun$ haloes with different feedback models. This is expected from the large difference in the resulting CGM phase structure which is shown in the top panel of Fig. \ref{fig:Phase_Evolution} and Fig. \ref{fig:Phase_Evolution_M11}. Therefore, C\thinspace\textsc{iv} can be used to constrain the nature of galactic winds. Indeed, \cite{Bordoloi+14} demonstrated the efficacy of this technique by comparing their observations to simulations that use two different feedback models. Our results in Fig. \ref{fig:OVI_HI_Column_Prof} qualitatively favor models with large mass loading $\eta$ and modest wind speeds. 

Observations of H\thinspace\textsc{i} around $z\sim0$ L$^*$ ($\Mh \approx 10^{12}~\Msun$) galaxies find typical column densities of $\gtrsim10^{16.5}$ cm$^{-2}$ at $\sim75$kpc \citep{Thom+12, Werk+14}, and $\sim10^{14.5}$ cm$^{-2}$ at $\sim150$ kpc \citep{Prochaska+11}. 
The H\thinspace\textsc{i} content of our simulations, shown in Figs. \ref{fig:OVI_HI_Column_Prof} and \ref{fig:OVI_HI_Column}, are below the observed values by an order of magnitude or more. This discrepancy is similar to what has been found in cosmological simulations \citep[e.g.,][]{Hummels+13}. Note that the N$_{\rm H\thinspace\textsc{i}}$ of the $10^{11}~\Msun$ haloes in the central $\sim60$kpc varies by up to two orders of magnitude depending on the feedback model{, whereas in the more massive halos the N$_{\rm H\thinspace\textsc{i}}$ is less sensitive to the strength of feedback \citep[see also][]{Rahmati+15}.}
Matching the observed column densities of cooler gas $\lesssim 3\times10^{4}$ K that is traced by H\thinspace\textsc{i} with hydrodynamic simulations is challenging because of the stringent spatial resolution requirements necessary to resolve this phase. 
Observations of the cold CGM at higher redshift, $z\sim1-2$, imply clump sizes ($\ell \sim N/n$) on the order of $\lesssim10$pc \citep[e.g.][]{ProchaskaHennawi09, Hennawi+15}.
There are some claims, however, that the $z\sim0$ COS-haloes observations point to much larger cold clump sizes of $\sim 10$kpc \citep{Werk+14}. This result is based on the assumption that all of the absorbers are at the same density. The best fit cold clump size in an alternative model, which assumes small, high density clouds are hierarchically nested within successively larger and less dense clouds, is on the order of 6 pc \citep{Stern+16}. The $\lesssim$ kpc resolution of our simulations at $\rvir$, which is better resolution than in the CGM of most cosmological simulations, is likely sufficient to resolve the warm gas traced by O\thinspace\textsc{vi} gas, but may be insufficient to fully capture the colder H\thinspace\textsc{i} gas. 

The right panel of Fig. \ref{fig:OVI_HI_Column} shows a typical example of the density weighted line-of-sight velocity in our simulations. The snapshot comes from the $10^{11.5}~\Msun$ halo with the high $\eta$ feedback model after 6 Gyr of evolution. In practice, we measure the density weighted mean of the absolute value of the line-of-sight velocity, which avoids cancellation and is comparable to the observed measurements of CGM absorption feature velocity offsets relative to their host galaxy. The line-of-sight velocities in our haloes peak near the center at $\sim 100$ km~s$^{-1}$ and drop to $\lesssim 10$ km~s$^{-1}$ near $\rvir$. This is reasonably consistent with observations (e.g., for O\thinspace\textsc{vi}, see Figure 2B of \citealt{Tumlinson+11}). In future work we plan to look at the kinematics of the halo gas traced by different ions for comparison to the recent analysis of the kinematics in the COS-Halos sample that shows distinct changes in line-widths at different halo masses as would be expected from our simulations (see Fig. \ref{fig:support}) \citep{Werk+16}.

\section{Discussion} \label{Discussion}

We have carried out a suite of idealized three-dimensional hydrodynamic simulations of the baryonic content of dark matter haloes with masses ranging from $10^{11}$ to $10^{12}~\Msun$ that include the effects of cooling, galactic winds driven by stellar feedback, and cosmological gas accretion. For each halo mass we consider, we adopted feedback models that have mass-loading factors, $\eta$, which bracket the expected range for star formation feedback ($\eta$ is the ratio of the galactic wind outflow rate to star formation rate; see equation (\ref{eq:eta})). This provides a controlled setup for understanding the impact of stellar feedback on the CGM. 

The standard paradigm in galaxy formation \citep{ReesOstriker+77,Silk77,Binney77, BirnboimDekel03}, which omitted feedback, identified a critical halo mass ($\sim10^{11.5}~\Msun$ relatively independent of $z$), above which the virial shock heated halo gas cools slowly and remains thermally supported for many dynamical times. Alternatively, in haloes below this critical mass the virial shock heated gas cools rapidly and thermal pressure alone is insufficient to support the gas. With our idealized simulations we have added the additional ingredient of stellar feedback without the complexity inherent to cosmological simulations that have also addressed this question. We show that feedback does not significantly alter the critical mass that delineates haloes with and without thermal pressure supported gaseous haloes (see Fig. \ref{fig:support}).

Our simulations demonstrate that the impact of feedback on the thermal structure and dynamics of the CGM differs above and below the critical halo mass. In more massive haloes, $\gtrsim 10^{11.5}~\Msun$, the state of the halo gas at large radii $\sim \rvir$ is relatively insensitive to the choice of feedback model. At small radii, near the central galaxy, halo gas is regulated by feedback. Feedback is triggered by accretion of cold gas which condenses out of the hotter ambient medium. With the low $\eta$ (i.e., high energy per unit mass) feedback model this condensation is correlated with epochs when $t_{\rm cool}/t_{\rm ff}~\lesssim~10$, which is indicative of the condensation being triggered by thermal instability (Fig. \ref{fig:tcool_tff_evolution}). The resulting heating stabilizes the hot gas against further condensation and reduces the star formation rate by a factor of $2-10$ relative to haloes without feedback (see Fig. \ref{fig:Mdots_Plot}). Above a certain minimum level of feedback (the actual value depends on the specific feedback parameterization) changes to the feedback efficiency have a minor impact on the global properties of the halo gas, the main difference being that less efficient feedback heats the CGM less effectively so the haloes spend more time with $t_{\rm cool} / t_{\rm ff} \lesssim 1-10$ (see Fig. \ref{fig:tcool_tff_evolution}). This is in accordance with the findings of similar studies targeting the group and cluster regime \citep[e.g.,][]{Sharma+12a} and cosmological simulations targeting galaxy mass halos \citep[e.g.,][]{vandeVoort+12,Rahmati+15} .

Below the critical halo mass, $\sim10^{11.5}$ $\Msun$, turbulence and bulk flows play a larger role in supporting halo gas (Fig. \ref{fig:support}). In this regime, feedback and its interaction with inflowing gas determines the properties of the CGM. Gas in these haloes is far from hydrostatic equilibrium.
Changes to the feedback efficiency lead to dramatic differences in the phase structure of the CGM (Figures \ref{fig:Phase_Evolution} and \ref{fig:Phase_Evolution_M11}). 
If the galactic wind is heavily mass loaded then much of the mass in the CGM is in the form of dense, cold, outgoing clumps surrounded by relatively sparse inflowing cold gas. The density and velocity of the wind ejecta in this case cause the wind shock cooling times to be very short leading to a paucity of warm, $\gtrsim 10^5$ K, gas. Alternatively, winds with mass loading $< 1$ with a similar total energetic efficiency lead to a CGM structure that is very different: a multiphase medium with appreciable mass between $\sim 10^4$~K and $10^6$~K (compare Fig. \ref{fig:Phase_Evolution_M11} with the top panel of Fig. \ref{fig:Phase_Evolution}) that is supported in part by turbulence and is threaded by dense gas flowing inwards along narrow channels (Figs. \ref{fig:M11_slices} and \ref{fig:support}). These two feedback models bracket the expected range of wind mass loading \citep[e.g.][]{Martin99, Veilleux+2005, Heckman+15, Muratov+15}, and wind velocities \citep[e.g.,][]{Heckman+00,Martin+05,Weiner+09,Rubin+11}.

An important difference between the CGM of different mass haloes is that the thermal instability triggered feedback regulation of hot halo gas {(condensation and feedback triggered when $t_{\rm cool} / t_{\rm ff}$ drops below $\sim 10$)} that successfully explains much of the structure and evolution of massive haloes ($>10^{12}~\Msun$) breaks down for halo masses $\lesssim10^{12}~\Msun$ (see Fig. \ref{fig:tcool_tff_evolution}). Therefore, the attempts of \cite{Voit+15b} to explain galaxy properties using this model may not be applicable at these lower halo masses. 

In Section \ref{Observations} we present a comparison of our simulations to some of the key results that have come out of quasar absorption-line observations of the CGM in the $z\sim0$ universe \citep[e.g.,][]{Tumlinson+11, Thom+12, Stocke+13, Werk+14,Borthakur+15}. In particular, we show the O\thinspace\textsc{vi}, C\thinspace\textsc{iv} and H\thinspace\textsc{i} content of our haloes, as well as a representative example of the density weighted line-of-sight velocity (Figures \ref{fig:OVI_HI_Column_Prof} and \ref{fig:OVI_HI_Column}).
The idealized nature of our simulations prevents us from making too detailed of a comparison to the observations. Bearing this in mind, the O\thinspace\textsc{vi} and C\thinspace\textsc{iv} column densities of our haloes are close to the observed values \citep{Tumlinson+11, Bordoloi+14}. Likewise, the density weighted line-of-sight velocities of our halo gas ($\sim 100$ km~s$^{-1}$) are similar to velocity offsets of the CGM absorption features relative to their host galaxy in background quasar spectra \citep[e.g.,][]{Tumlinson+11, Thom+12}.
Interestingly, we find a non-monotonic dependence of N$_{\rm O \thinspace\textsc{vi}}$ on halo mass, with the column densities peaking at $\sim10^{11.5}~\Msun$ haloes.
Several cosmological simulations have tried to reproduce these N$_{\rm O \thinspace\textsc{vi}}$ observations with varying degrees of success \citep[e.g.,][]{TepperGarcia+11,Hummels+13,Ford+15, Suresh+15b, Liang+16}.
The inclusion of additional physics, such as cosmic rays \citep{Salem+16} and non-equilibrium ionization \citep{Oppenheimer+16}, has improved the correspondence with the N$_{\rm O\thinspace\textsc{vi}}$ observations. The results from the latter study also indicate a possible strong dependence of the {O\thinspace\textsc{vi}} column density on halo mass, which is seen in our simulations. It is also worth noting that different feedback models lead to different N$_{\rm O\thinspace\textsc{vi}}$ and N$_{\rm C\thinspace\textsc{iv}}$ profiles, particularly at lower halo masses, which may enable observations of the CGM to constrain the nature of galactic winds. This is particularly promising in lower mass haloes, as probed e.g. by the COS-Dwarfs sample \citep{Bordoloi+14}.

The H\thinspace\textsc{i} content of our haloes (Fig. \ref{fig:OVI_HI_Column_Prof}) are well below the observed values \citep[e.g.,][]{Prochaska+11,Werk+14}. Although the neutral hydrogen column densities are under-predicted, the total hydrogen column densities in our simulations agree well with values implied by photoionization models of the COS-haloes observations \citep{Werk+14}. This implies that our haloes have roughly the right amount of total gas, but not enough gas that is dense enough for self-shielding to allow the neutral fraction to reach an appreciable value. At $z=0$ the density above which self-shielding becomes important is $n_{\rm SSh} \approx 3\times10^{-3}$cm$^{-3}$ \citep{Rahmati+13}. Fig. \ref{fig:PDF} shows that in our simulations the characteristic density is $n\sim3\times10^{-5}$ cm$^{-3}$ (at radii where neutral hydrogen is observed $\gtrsim 0.5\rvir$), which is well below $n_{\rm SSh}$. There are a few plausible explanations for why there is less sufficiently dense ($n\gtrsim n_{\rm SSh}$) gas at large radii in our haloes than is implied by the observations. The dense neutral hydrogen containing gas may reside in very small clumps ($\ell \lesssim 10$pc), as suggested by observations \citep[e.g.][]{ProchaskaHennawi09, Hennawi+15,Stern+16}. These small clumps may have formed as result of rapid cooling of galactic wind shocked material \citep[e.g.][]{Thompson+16}, or as a result of pressure confinement after condensing out of a hotter, thermally unstable background \citep[e.g.][]{McCourt+12, McCourt+16}. These small clumps would be unresolved in our simulations, so the gas would not be able to reach the large densities necessary to explain the large observed H\thinspace\textsc{i} columns. Alternatively, the H\thinspace\textsc{i} may be due to substructure in the dark matter halo, which is not included in our simulations. Satellite galaxies could provide the necessary binding energy to hold dense clouds together when pressure confinement is insufficient, and their winds could inject more dense gas into the CGM. Finally, if the large scale accretion proceeds along filaments the density of the inflowing gas may be high enough for there to be an appreciable amount of H\thinspace\textsc{i}. Note, however, that this cannot explain the ${\sim}1$ covering fraction of H\thinspace\textsc{i} absorbers with N$_{\rm H\thinspace\textsc{i} }>10^{15}\mathrm{~cm}^{-2}$ \citep{Tumlinson+13}. It is likely that all of these effects (and more), which are not included in our simulations, come into play in setting the amount of H\thinspace\textsc{i} in real haloes. It is important to stress that our idealized calculations are able to roughly reproduce the observed properties of C\thinspace\textsc{iv} and O\thinspace\textsc{vi} absorbers, suggesting that these ions may be better than H\thinspace\textsc{i} as probes of the overall impact of star formation feedback on the CGM.

The physical properties of the haloes we simulated were chosen to represent haloes in the $z=0$ universe, which allows us to compare to COS observations. However, our results are applicable to higher redshifts as well because of the weak redshift dependence of $t_{\rm cool} / t_{\rm ff}$ at the virial shock \citep{DekelBirnboim06}. This ratio determines whether a virial shock will remain hot and grow or cool rapidly and collapse. We have verified this argument by carrying out a suite of $z=2$ simulations, which showed the same qualitative behavior as their $z=0$ counterparts, for a given set of feedback parameters. Of course, it may well be that the efficiency of stellar and/or AGN feedback vary with redshift, due to changing gas fractions, black hole masses, metallicities, and other physical characteristics of the galaxies and haloes. If so, the CGM properties would vary as well. Although the redshift-independence of our results implies that for a given mass our results are valid for a range of redshifts, the numerical setup we adopted for this paper neglects to account for the growth of the underlying dark matter potential. Moreover our simulations use only the mean cosmic accretion rate and do not account for any scatter or evolution of this rate, which could have potentially important implications for the CGM structure \citep{McCourt+13} and for galaxy properties \citep{Feldmann+16}. In a future study we plan to include an evolving potential and explore variations to the accretion rate at fixed halo mass.

In a similar vein, our simulations neglect large scale filamentary accretion, which is the norm in galaxy formation. \cite{Nelson+13} argued that the long standing cold-mode--hot-mode dichotomy \citep[e.g.,][] {Keres+05} is partially a numerical artifact arising from deficiencies in smooth-particle hydrodynamics (SPH) codes that lead to thinner, denser filaments, and heating due to spurious dissipation of turbulent motions at large scales. The lack of filaments in our current simulations precludes us from addressing this concern. However, the controlled environment our setup affords is ideal for studying how the properties of gaseous haloes vary with filament properties and stellar feedback. This is particularly important given the numerical subtleties involved in capturing mixing via the Kelvin-Helmholtz instability \citep{Lecoanet+15} and its supersonic variants \citep{Belyaev+12, Mandelker2016}. In a future study we plan to repeat a similar set of simulations as those in this paper, but with filamentary accretion. Magnetic fields, anisotropic conduction, and cosmic rays may also play an important role in the evolution and phase structure of the CGM \citep[e.g.,][]{Booth+13,McCourt+15}. 

\section*{Acknowledgements}
We thank Claude-Andr\'e Faucher-Gigu\`ere, Freeke van de Voort, Robert Feldmann, Joe Hennawi, Prateek Sharma, Joel Bregman, Zachary Hafen, and Jonathan Stern for useful conversations, and the anonymous referee for useful comments that improved this work.
Additionally, we thank Freeke van de Voort for providing the \textsc{cloudy} results on O\thinspace\textsc{vi} and C\thinspace\textsc{iv} fractions in PIE used in Figures \ref{fig:OVI_HI_Column_Prof} and \ref{fig:OVI_HI_Column}.
This work was supported in part by NASA ATP grant 12-APT12-0183 and a Simons Investigator award from the Simons Foundation to EQ. DF was supported by the NSF GRFP under Grant \# DGE 1106400. MM was supported by NASA grant \# NNX15AK81G. TAT is supported by NSF Grant \#1516967. TAT and EQ thank the Simons Foundation and organizers Juna Kollmeier and Andrew Benson for support for the {\it Galactic Winds: Beyond Phenomenology} symposium series, where this work germinated.

This research used the Savio computational cluster resource provided by the Berkeley Research Computing program at the University of California, Berkeley (supported by the UC Berkeley Chancellor, Vice Chancellor of Research, and Office of the CIO). In addition, this work used the Extreme Science and Engineering Discovery Environment (XSEDE), which is supported by National Science Foundation grant number ACI-1053575. Much of our analysis was performed using the publicly available data analysis software package \texttt{yt} \citep{yt}.

\begin{appendix} 
\section{Resolution study}\label{Appendix}

\begin{figure}
\hspace*{-0.5cm}   
\vspace*{-0.75cm}   
\includegraphics[width=0.5\textwidth]{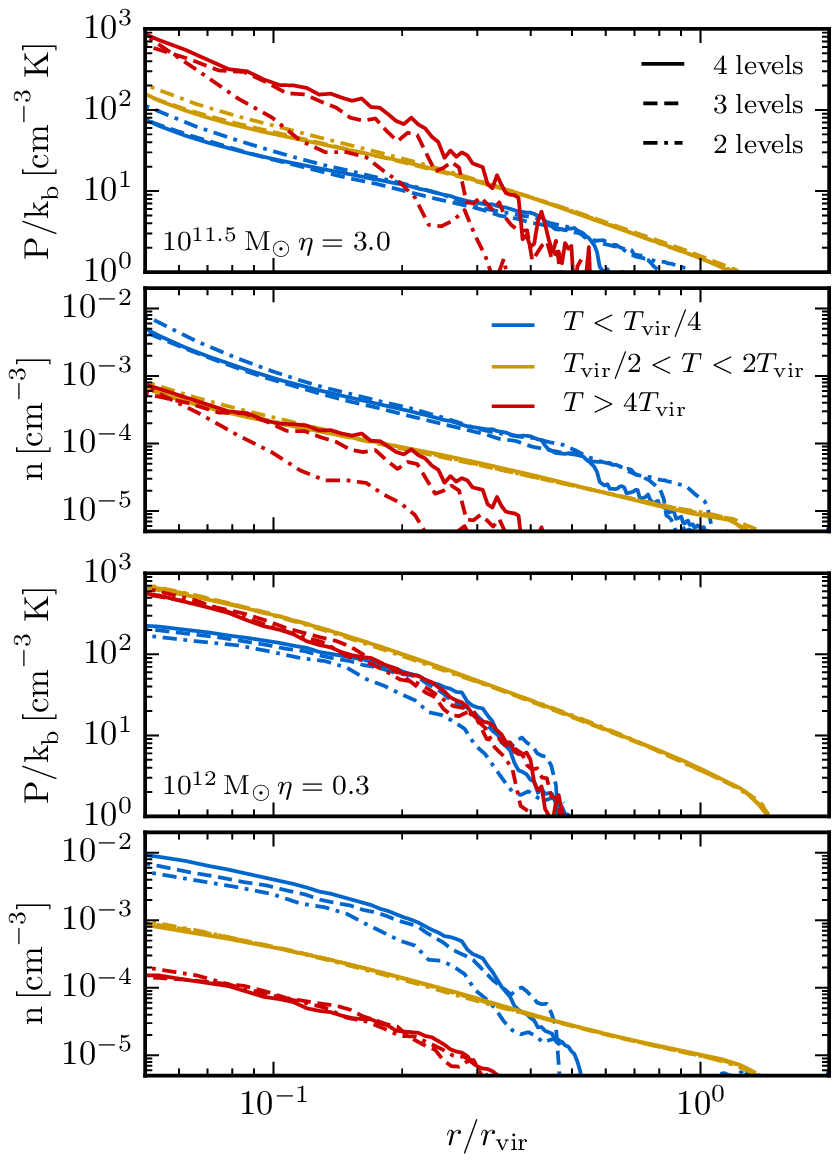}
\caption{The time averaged (from 3 to 9 Gyr) pressure and number density profiles of gas in three temperature bins that are delineated relative to the haloes' virial temperatures, for the high $\eta$ $10^{11.5}~\Msun$ (top) and low $\eta$ $10^{12}~\Msun$ haloes (bottom). The hot gas with $T>4T_{\rm vir}$, virial gas with $T_\mathrm{vir}/2<T<2T_\mathrm{vir}$, and cold gas with $T<T_\mathrm{vir}/4$ are shown in red, gold, and blue, respectively. We show the profiles from the four mesh refinement level (solid lines), fiducial resolution, three mesh refinement level (dashed lines), and low resolution, two mesh refinement level (dot-dashed lines) simulations. The profiles change from low to fiducial resolution, but are similar at fiducial and high resolution, indicating that our simulations are close to converged.} \label{fig:Pressure_Equilibrium_convergence}
\vspace*{-0.4cm}   
\end{figure}

\begin{figure}
\hspace*{-0.25cm}   
\vspace*{-0.2cm}   
\includegraphics[width=0.525\textwidth]{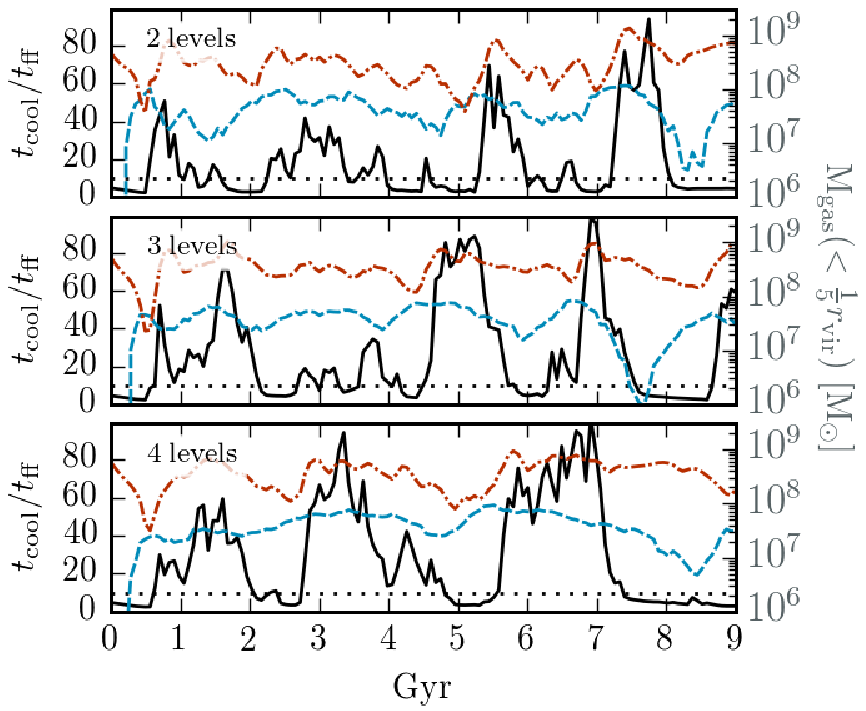}
\caption{ Spherical shell averaged $\tcool/\tff$ evolution at $r=3\rgal=0.075\rvir$ (left vertical axis) relative to the cold (T $<10^{4.5}$K; blue dashed line) and hot (T $>10^{6}$K; red dot-dashed line) gas mass contained within 0.2 $\rvir$ (right vertical axis) for the low $\eta$ $10^{12}~\Msun$ halo simulation with resolution increasing from top to bottom. A thin dotted line is drawn at  $\tcool/\tff=10$, the value below which hot halo gas is predicted to generate significant multiphase gas via thermal instability. The increase of cold gas mass when $\tcool/\tff<10$ is a strong indication of thermal instability. The thermal instability cycles are qualitatively similar, but quantitatively different because of the difficulty in resolving the small dense blobs created by thermal instability (e.g., Fig. \ref{fig:M12_projection}).} \label{fig:tcool_tff_convergence}
\vspace*{-0.2cm}   
\end{figure}

\begin{figure*}
\hspace*{-0.5cm}   
\vspace*{-0.5cm}   
\includegraphics[width=1.1\textwidth]{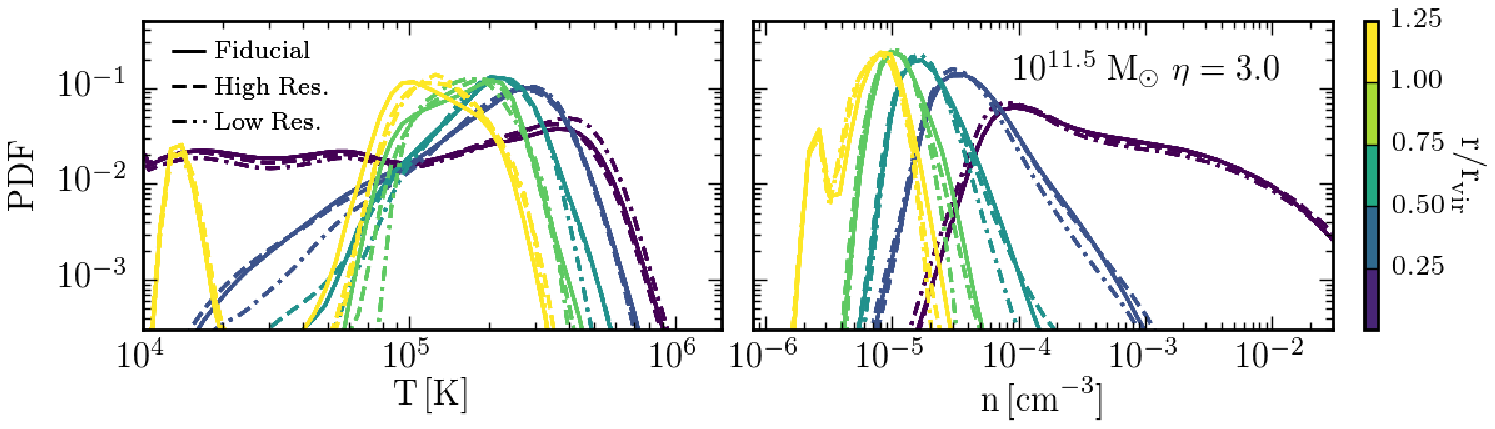}
\caption{ Mass weighted probability distribution as a function of temperature (left) and number density (right) for the high $\eta$ $10^{11.5}~\Msun$ halo. The different colors correspond to different spherical annuli ranging from the center (excluding the galaxy) to $1.25~\rvir$ in steps of $0.25~\rvir$. The structure of the haloes in the high resolution (two level 512$^3$) simulation agrees very well with the structure from the fiducial resolution (three level 256$^3$) and the low resolution (four level $128^3$) simulations.} \label{fig:PDF_convergence}
\vspace*{-0.35cm}   
\end{figure*}

We test the numerical convergence of our results with two separate tests. In the primary test we vary the number of levels of mesh refinements of two of our simulations, which allows us to assess how the properties of the halo cores vary with resolution. The fiducial resolution of our simulations is achieved using three, nested $256^3$ levels (base grid, and two nested refined levels). This gives a central spatial resolution of 228 cells per $\rvir$, or $\Delta x = 1.4$~kpc~$(\Mh/10^{12} \Msun)^{1/3}$.
We re-simulated the low $\eta$ $10^{12}~\Msun$ and high $\eta$ $10^{11.5}~\Msun$ haloes using two and four levels, which halved and doubled the central resolution, respectively. 

Fig. \ref{fig:Pressure_Equilibrium_convergence} shows the number density and pressure profiles of gas in three distinct temperature regimes of these two haloes (compare with Fig. \ref{fig:n_profile_Tbins}).
The profiles of the cold gas ($T<T_\mathrm{vir}/4$) are in blue, the warm, virialized gas ($T_\mathrm{vir}/2<T<2T_\mathrm{vir}$) in gold, and the hot gas ($T>4T_{\rm vir}$) in red. The agreement between the profiles for the different resolution simulations of the high $\eta$ $10^{11.5}~\Msun$ halo (top two panels) is very good in the cold and warm gas. However, the central densities and pressures of the hot gas in the high resolution (solid lines) simulations exceeds that of the fiducial resolution (dashed lines) and low resolution (dot-dashed lines) simulations. Reassuringly, the differences between the high and fiducial resolution simulations are much smaller than the differences between the fiducial and low resolution simulations, indicating that our simulations are approaching convergence in this hottest phase. This hot gas in the center of these haloes is heated by shocks between the wind material and the ambient/inflowing material. 

The temperature dependence of the resolution sensitivity is reversed in the low $\eta$ $10^{12}~\Msun$ halo (bottom two panels). In this case, with different resolutions the hot and warm gas profiles agree very well, while the cold gas pressures and densities increase with resolution. Again, however, the differences decrease as resolution increases. The gas in this cold temperature range is primarily in small pressure confined clumps (e.g. Fig. \ref{fig:M12_projection}). The size of these clumps should be determined either by thermal conduction, which would require resolving the Field length ($\lambda_F\lesssim 0.1 ~\mathrm{pc}\ll \Delta x$ for $10^{-2}$~cm$^{-3}$ gas at $10^4$ K), or should be roughly the size where the sound crossing time equals the cooling time ($\sim c_s t_{\rm cool} \lesssim 3$pc). These clumps are therefore unlikely to be resolved in our simulations (or any other full halo simulations). 

Fig. \ref{fig:tcool_tff_convergence} shows a convergence plot for the evolution of the ratio of the cooling time $\tcool$ of the hot gas at $3\rgal=0.075\rvir$ to its free fall time $\tff$ along with the cold and hot gas content of the inner 0.2$\rvir$ for the low $\eta$ $10^{12}~\Msun$ halo at different resolutions (the middle panel is the same as the top panel of Fig. \ref{fig:tcool_tff_evolution}). The increase of cold gas only when $\tcool/\tff < 10$ and the corresponding increase in hot gas mass and $\tcool/\tff$ immediately after is a strong indicator of thermal instability triggered feedback regulation. 
As resolution increases there are fewer of these cycles over a fixed length of time. 
Additionally, at low resolution the increases in the cold gas mass are more abrupt and the duration of the $\tcool/\tff > 10$ phase is shorter than in the higher resolution simulations.
Although the precise details of this regulation and its indicators change with resolution, the overall behavior is similar at the resolutions we considered. 

The second numerical convergence test we performed was aimed at assessing how our results depend on the resolution at large radii ($\gtrsim 0.5 \rvir$). To do so we simulated the low $\eta$ $10^{11}~\Msun$, high $\eta$ $10^{11.5}~\Msun$, and low $\eta$ $10^{12}~\Msun$ haloes with the same central resolution but differing base grid resolution. For these tests the high resolution simulations have two nested $512^3$ grids and the low resolution simulations have four nested $128^3$ grids, compared to the three nested $256^3$ grids used in our fiducial resolution simulations. 

Fig. \ref{fig:PDF_convergence} shows the mass-weighted probability distribution as a function of temperature and number density in five radial bins for the high $\eta$ $10^{11.5}~\Msun$ halo. The low $\eta$ $10^{11}~\Msun$ and low $\eta$ $10^{12}~\Msun$ haloes show a similar degree of convergence so they are not shown. At all three halo masses the agreement, especially at large radii, is very good, which demonstrates that the large scale properties of our haloes are well resolved in our simulations.

\bsp

\end{appendix}

\bibliographystyle{mnras}
\bibliography{Galaxyhalo}

\begin{thebibliography}{}
\makeatletter
\relax
\def\mn@urlcharsother{\let\do\@makeother \do\$\do\&\do\#\do\^\do\_\do\%\do\~}
\def\mn@doi{\begingroup\mn@urlcharsother \@ifnextchar [ {\mn@doi@}
  {\mn@doi@[]}}
\def\mn@doi@[#1]#2{\def\@tempa{#1}\ifx\@tempa\@empty \href
  {http://dx.doi.org/#2} {doi:#2}\else \href {http://dx.doi.org/#2} {#1}\fi
  \endgroup}
\def\mn@eprint#1#2{\mn@eprint@#1:#2::\@nil}
\def\mn@eprint@arXiv#1{\href {http://arxiv.org/abs/#1} {{\tt arXiv:#1}}}
\def\mn@eprint@dblp#1{\href {http://dblp.uni-trier.de/rec/bibtex/#1.xml}
  {dblp:#1}}
\def\mn@eprint@#1:#2:#3:#4\@nil{\def\@tempa {#1}\def\@tempb {#2}\def\@tempc
  {#3}\ifx \@tempc \@empty \let \@tempc \@tempb \let \@tempb \@tempa \fi \ifx
  \@tempb \@empty \def\@tempb {arXiv}\fi \@ifundefined
  {mn@eprint@\@tempb}{\@tempb:\@tempc}{\expandafter \expandafter \csname
  mn@eprint@\@tempb\endcsname \expandafter{\@tempc}}}

\bibitem[\protect\citeauthoryear{{Anderson} \& {Bregman}}{{Anderson} \&
  {Bregman}}{2011}]{AndersonBregman+11}
{Anderson} M.~E.,  {Bregman} J.~N.,  2011, \mn@doi [ApJ]
  {10.1088/0004-637X/737/1/22}, \href
  {http://adsabs.harvard.edu/abs/2011ApJ...737...22A} {737, 22}

\bibitem[\protect\citeauthoryear{{Balbus}}{{Balbus}}{2001}]{Balbus01}
{Balbus} S.~A.,  2001, \mn@doi [ApJ] {10.1086/323875}, \href
  {http://adsabs.harvard.edu/abs/2001ApJ...562..909B} {562, 909}

\bibitem[\protect\citeauthoryear{{Behroozi}, {Conroy}  \&
  {Wechsler}}{{Behroozi} et~al.}{2010}]{Behroozi+10}
{Behroozi} P.~S.,  {Conroy} C.,   {Wechsler} R.~H.,  2010, \mn@doi [ApJ]
  {10.1088/0004-637X/717/1/379}, \href
  {http://adsabs.harvard.edu/abs/2010ApJ...717..379B} {717, 379}

\bibitem[\protect\citeauthoryear{{Belyaev} \& {Rafikov}}{{Belyaev} \&
  {Rafikov}}{2012}]{Belyaev+12}
{Belyaev} M.~A.,  {Rafikov} R.~R.,  2012, \mn@doi [ApJ]
  {10.1088/0004-637X/752/2/115}, \href
  {http://adsabs.harvard.edu/abs/2012ApJ...752..115B} {752, 115}

\bibitem[\protect\citeauthoryear{{Bett} \& {Frenk}}{{Bett} \&
  {Frenk}}{2012}]{BettFrenk}
{Bett} P.~E.,  {Frenk} C.~S.,  2012, \mn@doi [MNRAS]
  {10.1111/j.1365-2966.2011.20275.x}, \href
  {http://adsabs.harvard.edu/abs/2012MNRAS.420.3324B} {420, 3324}

\bibitem[\protect\citeauthoryear{{Binney}}{{Binney}}{1977}]{Binney77}
{Binney} J.,  1977, \mn@doi [ApJ] {10.1086/155378}, \href
  {http://adsabs.harvard.edu/abs/1977ApJ...215..483B} {215, 483}

\bibitem[\protect\citeauthoryear{{Birnboim} \& {Dekel}}{{Birnboim} \&
  {Dekel}}{2003}]{BirnboimDekel03}
{Birnboim} Y.,  {Dekel} A.,  2003, \mn@doi [MNRAS]
  {10.1046/j.1365-8711.2003.06955.x}, 345, 349

\bibitem[\protect\citeauthoryear{{Book}, {Brooks}, {Peter}, {Benson}  \&
  {Governato}}{{Book} et~al.}{2011}]{Book+11}
{Book} L.~G.,  {Brooks} A.,  {Peter} A.~H.~G.,  {Benson} A.~J.,   {Governato}
  F.,  2011, \mn@doi [MNRAS] {10.1111/j.1365-2966.2010.17824.x}, \href
  {http://adsabs.harvard.edu/abs/2011MNRAS.411.1963B} {411, 1963}

\bibitem[\protect\citeauthoryear{Booth, Agertz, Kravtsov  \& Gnedin}{Booth
  et~al.}{2013}]{Booth+13}
Booth C.~M.,  Agertz O.,  Kravtsov A.~V.,   Gnedin N.~Y.,  2013, ApJL, 777, L16

\bibitem[\protect\citeauthoryear{{Bordoloi} et~al.,}{{Bordoloi}
  et~al.}{2014}]{Bordoloi+14}
{Bordoloi} R.,  et~al., 2014, \mn@doi [ApJ] {10.1088/0004-637X/796/2/136},
  \href {http://adsabs.harvard.edu/abs/2014ApJ...796..136B} {796, 136}

\bibitem[\protect\citeauthoryear{{Borthakur} et~al.,}{{Borthakur}
  et~al.}{2015}]{Borthakur+15}
{Borthakur} S.,  et~al., 2015, \mn@doi [ApJ] {10.1088/0004-637X/813/1/46}, 813,
  46

\bibitem[\protect\citeauthoryear{{Dekel} \& {Birnboim}}{{Dekel} \&
  {Birnboim}}{2006}]{DekelBirnboim06}
{Dekel} A.,  {Birnboim} Y.,  2006, \mn@doi [MNRAS]
  {10.1111/j.1365-2966.2006.10145.x}, \href
  {http://adsabs.harvard.edu/abs/2006MNRAS.368....2D} {368, 2}

\bibitem[\protect\citeauthoryear{{Dekel} et~al.,}{{Dekel}
  et~al.}{2009}]{Dekel+09}
{Dekel} A.,  et~al., 2009, \mn@doi [Nature] {10.1038/nature07648}, \href
  {http://adsabs.harvard.edu/abs/2009Natur.457..451D} {457, 451}

\bibitem[\protect\citeauthoryear{{Fang}, {Bullock}  \& {Boylan-Kolchin}}{{Fang}
  et~al.}{2013}]{Fang+13}
{Fang} T.,  {Bullock} J.,   {Boylan-Kolchin} M.,  2013, \mn@doi [ApJ]
  {10.1088/0004-637X/762/1/20}, \href
  {http://adsabs.harvard.edu/abs/2013ApJ...762...20F} {762, 20}

\bibitem[\protect\citeauthoryear{{Faucher-Gigu{\`e}re}, {Kere{\v s}}  \&
  {Ma}}{{Faucher-Gigu{\`e}re} et~al.}{2011}]{FaucherGiguere+11}
{Faucher-Gigu{\`e}re} C.-A.,  {Kere{\v s}} D.,   {Ma} C.-P.,  2011, \mn@doi
  [MNRAS] {10.1111/j.1365-2966.2011.19457.x}, \href
  {http://adsabs.harvard.edu/abs/2011MNRAS.417.2982F} {417, 2982}

\bibitem[\protect\citeauthoryear{{Faucher-Gigu{\`e}re}, {Hopkins}, {Kere{\v
  s}}, {Muratov}, {Quataert}  \& {Murray}}{{Faucher-Gigu{\`e}re}
  et~al.}{2015}]{CAFG+15}
{Faucher-Gigu{\`e}re} C.-A.,  {Hopkins} P.~F.,  {Kere{\v s}} D.,  {Muratov}
  A.~L.,  {Quataert} E.,   {Murray} N.,  2015, \mn@doi [MNRAS]
  {10.1093/mnras/stv336}, \href
  {http://adsabs.harvard.edu/abs/2015MNRAS.449..987F} {449, 987}

\bibitem[\protect\citeauthoryear{{Faucher-Gigu{\`e}re}, {Feldmann}, {Quataert},
  {Keres}, {Hopkins}  \& {Murray}}{{Faucher-Gigu{\`e}re}
  et~al.}{2016}]{CAFG+16}
{Faucher-Gigu{\`e}re} C.-A.,  {Feldmann} R.,  {Quataert} E.,  {Keres} D.,
  {Hopkins} P.~F.,   {Murray} N.,  2016, preprint, \href
  {http://adsabs.harvard.edu/abs/2016arXiv160107188F} {} (\mn@eprint {arXiv}
  {1601.07188})

\bibitem[\protect\citeauthoryear{{Feldmann}, {Hopkins}, {Quataert},
  {Faucher-Giguere}  \& {Keres}}{{Feldmann} et~al.}{2016}]{Feldmann+16}
{Feldmann} R.,  {Hopkins} P.~F.,  {Quataert} E.,  {Faucher-Giguere} C.-A.,
  {Keres} D.,  2016, preprint, \href
  {http://adsabs.harvard.edu/abs/2016arXiv160104704F} {} (\mn@eprint {arXiv}
  {1601.04704})

\bibitem[\protect\citeauthoryear{{Ferland}, {Korista}, {Verner}, {Ferguson},
  {Kingdon}  \& {Verner}}{{Ferland} et~al.}{1998}]{CLOUDY}
{Ferland} G.~J.,  {Korista} K.~T.,  {Verner} D.~A.,  {Ferguson} J.~W.,
  {Kingdon} J.~B.,   {Verner} E.~M.,  1998, \mn@doi [PASP] {10.1086/316190},
  \href {http://adsabs.harvard.edu/abs/1998PASP..110..761F} {110, 761}

\bibitem[\protect\citeauthoryear{{Ford}, {Oppenheimer}, {Dav{\'e}}, {Katz},
  {Kollmeier}  \& {Weinberg}}{{Ford} et~al.}{2013}]{Ford+13}
{Ford} A.~B.,  {Oppenheimer} B.~D.,  {Dav{\'e}} R.,  {Katz} N.,  {Kollmeier}
  J.~A.,   {Weinberg} D.~H.,  2013, \mn@doi [MNRAS] {10.1093/mnras/stt393},
  \href {http://adsabs.harvard.edu/abs/2013MNRAS.432...89F} {432, 89}

\bibitem[\protect\citeauthoryear{{Ford} et~al.,}{{Ford} et~al.}{2015}]{Ford+15}
{Ford} A.~B.,  et~al., 2015, preprint, \href
  {http://adsabs.harvard.edu/abs/2015arXiv150302084F} {} (\mn@eprint {arXiv}
  {1503.02084})

\bibitem[\protect\citeauthoryear{{Forman}, {Jones}  \& {Tucker}}{{Forman}
  et~al.}{1985}]{Forman+85}
{Forman} W.,  {Jones} C.,   {Tucker} W.,  1985, \mn@doi [ApJ] {10.1086/163218},
  \href {http://adsabs.harvard.edu/abs/1985ApJ...293..102F} {293, 102}

\bibitem[\protect\citeauthoryear{{Gardiner} \& {Stone}}{{Gardiner} \&
  {Stone}}{2008}]{Gardiner+08}
{Gardiner} T.~A.,  {Stone} J.~M.,  2008, \mn@doi [Journal of Computational
  Physics] {10.1016/j.jcp.2007.12.017}, 227, 4123

\bibitem[\protect\citeauthoryear{{Gaspari}, {Ruszkowski}  \&
  {Sharma}}{{Gaspari} et~al.}{2012}]{Gaspari+12}
{Gaspari} M.,  {Ruszkowski} M.,   {Sharma} P.,  2012, \mn@doi [ApJ]
  {10.1088/0004-637X/746/1/94}, \href
  {http://adsabs.harvard.edu/abs/2012ApJ...746...94G} {746, 94}

\bibitem[\protect\citeauthoryear{{Gray} \& {Scannapieco}}{{Gray} \&
  {Scannapieco}}{2016}]{Gray+16}
{Gray} W.~J.,  {Scannapieco} E.,  2016, \mn@doi [ApJ]
  {10.3847/0004-637X/818/2/198}, \href
  {http://adsabs.harvard.edu/abs/2016ApJ...818..198G} {818, 198}

\bibitem[\protect\citeauthoryear{{Gupta}, {Mathur}, {Krongold}, {Nicastro}  \&
  {Galeazzi}}{{Gupta} et~al.}{2012}]{Gupta+12}
{Gupta} A.,  {Mathur} S.,  {Krongold} Y.,  {Nicastro} F.,   {Galeazzi} M.,
  2012, \mn@doi [ApJL] {10.1088/2041-8205/756/1/L8}, \href
  {http://adsabs.harvard.edu/abs/2012ApJ...756L...8G} {756, L8}

\bibitem[\protect\citeauthoryear{{Haardt} \& {Madau}}{{Haardt} \&
  {Madau}}{2001}]{HaardtMadau01}
{Haardt} F.,  {Madau} P.,  2001, in {Neumann} D.~M.,  {Tran} J.~T.~V.,  eds,
  Clusters of Galaxies and the High Redshift Universe Observed in X-rays.
  (\mn@eprint {} {astro-ph/0106018})

\bibitem[\protect\citeauthoryear{{Heckman}, {Lehnert}, {Strickland}  \&
  {Armus}}{{Heckman} et~al.}{2000}]{Heckman+00}
{Heckman} T.~M.,  {Lehnert} M.~D.,  {Strickland} D.~K.,   {Armus} L.,  2000,
  \mn@doi [ApJs] {10.1086/313421}, \href
  {http://adsabs.harvard.edu/abs/2000ApJS..129..493H} {129, 493}

\bibitem[\protect\citeauthoryear{{Heckman}, {Alexandroff}, {Borthakur},
  {Overzier}  \& {Leitherer}}{{Heckman} et~al.}{2015}]{Heckman+15}
{Heckman} T.~M.,  {Alexandroff} R.~M.,  {Borthakur} S.,  {Overzier} R.,
  {Leitherer} C.,  2015, \mn@doi [ApJ] {10.1088/0004-637X/809/2/147}, 809, 147

\bibitem[\protect\citeauthoryear{{Hennawi}, {Prochaska}, {Cantalupo}  \&
  {Arrigoni-Battaia}}{{Hennawi} et~al.}{2015}]{Hennawi+15}
{Hennawi} J.~F.,  {Prochaska} J.~X.,  {Cantalupo} S.,   {Arrigoni-Battaia} F.,
  2015, \mn@doi [Science] {10.1126/science.aaa5397}, \href
  {http://adsabs.harvard.edu/abs/2015Sci...348..779H} {348, 779}

\bibitem[\protect\citeauthoryear{{Hummels}, {Bryan}, {Smith}  \&
  {Turk}}{{Hummels} et~al.}{2013}]{Hummels+13}
{Hummels} C.~B.,  {Bryan} G.~L.,  {Smith} B.~D.,   {Turk} M.~J.,  2013, \mn@doi
  [MNRAS] {10.1093/mnras/sts702}, \href
  {http://adsabs.harvard.edu/abs/2013MNRAS.430.1548H} {430, 1548}

\bibitem[\protect\citeauthoryear{{Kere{\v s}}, {Katz}, {Weinberg}  \&
  {Dav{\'e}}}{{Kere{\v s}} et~al.}{2005}]{Keres+05}
{Kere{\v s}} D.,  {Katz} N.,  {Weinberg} D.~H.,   {Dav{\'e}} R.,  2005, \mn@doi
  [MNRAS] {10.1111/j.1365-2966.2005.09451.x}, \href
  {http://adsabs.harvard.edu/abs/2005MNRAS.363....2K} {363, 2}

\bibitem[\protect\citeauthoryear{{Kere{\v s}}, {Vogelsberger}, {Sijacki},
  {Springel}  \& {Hernquist}}{{Kere{\v s}} et~al.}{2012}]{Keres+12}
{Kere{\v s}} D.,  {Vogelsberger} M.,  {Sijacki} D.,  {Springel} V.,
  {Hernquist} L.,  2012, \mn@doi [MNRAS] {10.1111/j.1365-2966.2012.21548.x},
  \href {http://adsabs.harvard.edu/abs/2012MNRAS.425.2027K} {425, 2027}

\bibitem[\protect\citeauthoryear{{Kunz}}{{Kunz}}{2011}]{Kunz11}
{Kunz} M.~W.,  2011, \mn@doi [MNRAS] {10.1111/j.1365-2966.2011.19303.x}, \href
  {http://adsabs.harvard.edu/abs/2011MNRAS.417..602K} {417, 602}

\bibitem[\protect\citeauthoryear{{Lecoanet} et~al.,}{{Lecoanet}
  et~al.}{2015}]{Lecoanet+15}
{Lecoanet} D.,  et~al., 2015, preprint (\mn@eprint {arXiv} {1509.03630})

\bibitem[\protect\citeauthoryear{{Li}, {Bryan}, {Ruszkowski}, {Voit}, {O'Shea}
  \& {Donahue}}{{Li} et~al.}{2015}]{Li+15}
{Li} Y.,  {Bryan} G.~L.,  {Ruszkowski} M.,  {Voit} G.~M.,  {O'Shea} B.~W.,
  {Donahue} M.,  2015, \mn@doi [ApJ] {10.1088/0004-637X/811/2/73}, 811, 73

\bibitem[\protect\citeauthoryear{{Liang}, {Kravtsov}  \& {Agertz}}{{Liang}
  et~al.}{2016}]{Liang+16}
{Liang} C.~J.,  {Kravtsov} A.~V.,   {Agertz} O.,  2016, \mn@doi [MNRAS]
  {10.1093/mnras/stw375}, \href
  {http://adsabs.harvard.edu/abs/2016MNRAS.458.1164L} {458, 1164}

\bibitem[\protect\citeauthoryear{{Maller} \& {Bullock}}{{Maller} \&
  {Bullock}}{2004}]{MallerBullock}
{Maller} A.~H.,  {Bullock} J.~S.,  2004, \mn@doi [MNRAS]
  {10.1111/j.1365-2966.2004.08349.x}, 355, 694

\bibitem[\protect\citeauthoryear{{Mandelker}, {Padnos}, {Dekel}, {Birnboim},
  {Burkert}, {Krumholz}  \& {Steinberg}}{{Mandelker}
  et~al.}{2016}]{Mandelker2016}
{Mandelker} N.,  {Padnos} D.,  {Dekel} A.,  {Birnboim} Y.,  {Burkert} A.,
  {Krumholz} M.~R.,   {Steinberg} E.,  2016, \mn@doi [MNRAS]
  {10.1093/mnras/stw2267}, \href
  {http://adsabs.harvard.edu/abs/2016MNRAS.tmp.1374M} {}

\bibitem[\protect\citeauthoryear{{Marasco}, {Debattista}, {Fraternali}, {van
  der Hulst}, {Wadsley}, {Quinn}  \& {Ro{\v s}kar}}{{Marasco}
  et~al.}{2015}]{Marasco+15}
{Marasco} A.,  {Debattista} V.~P.,  {Fraternali} F.,  {van der Hulst} T.,
  {Wadsley} J.,  {Quinn} T.,   {Ro{\v s}kar} R.,  2015, \mn@doi [MNRAS]
  {10.1093/mnras/stv1240}, \href
  {http://adsabs.harvard.edu/abs/2015MNRAS.451.4223M} {451, 4223}

\bibitem[\protect\citeauthoryear{{Martin}}{{Martin}}{1999}]{Martin99}
{Martin} C.~L.,  1999, \mn@doi [ApJ] {10.1086/306863}, \href
  {http://adsabs.harvard.edu/abs/1999ApJ...513..156M} {513, 156}

\bibitem[\protect\citeauthoryear{{Martin}}{{Martin}}{2005}]{Martin+05}
{Martin} C.~L.,  2005, \mn@doi [ApJ] {10.1086/427277}, \href
  {http://adsabs.harvard.edu/abs/2005ApJ...621..227M} {621, 227}

\bibitem[\protect\citeauthoryear{{McBride}, {Fakhouri}  \& {Ma}}{{McBride}
  et~al.}{2009}]{Mcbride+09}
{McBride} J.,  {Fakhouri} O.,   {Ma} C.-P.,  2009, \mn@doi [MNRAS]
  {10.1111/j.1365-2966.2009.15329.x}, 398, 1858

\bibitem[\protect\citeauthoryear{{McCourt}, {Parrish}, {Sharma}  \&
  {Quataert}}{{McCourt} et~al.}{2011}]{McCourt+11}
{McCourt} M.,  {Parrish} I.~J.,  {Sharma} P.,   {Quataert} E.,  2011, \mn@doi
  [MNRAS] {10.1111/j.1365-2966.2011.18216.x}, \href
  {http://adsabs.harvard.edu/abs/2011MNRAS.413.1295M} {413, 1295}

\bibitem[\protect\citeauthoryear{{McCourt}, {Sharma}, {Quataert}  \&
  {Parrish}}{{McCourt} et~al.}{2012}]{McCourt+12}
{McCourt} M.,  {Sharma} P.,  {Quataert} E.,   {Parrish} I.~J.,  2012, \mn@doi
  [MNRAS] {10.1111/j.1365-2966.2011.19972.x}, \href
  {http://adsabs.harvard.edu/abs/2012MNRAS.419.3319M} {419, 3319}

\bibitem[\protect\citeauthoryear{McCourt, Quataert  \& Parrish}{McCourt
  et~al.}{2013}]{McCourt+13}
McCourt M.,  Quataert E.,   Parrish I.~J.,  2013, Monthly Notices of the Royal
  Astronomical Society, 432, 404

\bibitem[\protect\citeauthoryear{{McCourt}, {O'Leary}, {Madigan}  \&
  {Quataert}}{{McCourt} et~al.}{2015}]{McCourt+15}
{McCourt} M.,  {O'Leary} R.~M.,  {Madigan} A.-M.,   {Quataert} E.,  2015,
  \mn@doi [MNRAS] {10.1093/mnras/stv355}, \href
  {http://adsabs.harvard.edu/abs/2015MNRAS.449....2M} {449, 2}

\bibitem[\protect\citeauthoryear{{McCourt}, {Oh}, {O'Leary}  \&
  {Madigan}}{{McCourt} et~al.}{2016}]{McCourt+16}
{McCourt} M.,  {Oh} S.~P.,  {O'Leary} R.~M.,   {Madigan} A.-M.,  2016,
  preprint, \href {http://adsabs.harvard.edu/abs/2016arXiv161001164M} {}
  (\mn@eprint {arXiv} {1610.01164})

\bibitem[\protect\citeauthoryear{{McNamara} \& {Nulsen}}{{McNamara} \&
  {Nulsen}}{2007}]{McNamaraNulsen2007}
{McNamara} B.~R.,  {Nulsen} P.~E.~J.,  2007, \mn@doi [AR\&A]
  {10.1146/annurev.astro.45.051806.110625}, \href
  {http://adsabs.harvard.edu/abs/2007ARA%26A..45..117M} {45, 117}

\bibitem[\protect\citeauthoryear{{Miller} \& {Bregman}}{{Miller} \&
  {Bregman}}{2015}]{Miller+15}
{Miller} M.~J.,  {Bregman} J.~N.,  2015, \mn@doi [ApJ]
  {10.1088/0004-637X/800/1/14}, \href
  {http://adsabs.harvard.edu/abs/2015ApJ...800...14M} {800, 14}

\bibitem[\protect\citeauthoryear{{Miller}, {Hodges-Kluck}  \&
  {Bregman}}{{Miller} et~al.}{2016}]{Miller+16}
{Miller} M.~J.,  {Hodges-Kluck} E.~J.,   {Bregman} J.~N.,  2016, \mn@doi [ApJ]
  {10.3847/0004-637X/818/2/112}, \href
  {http://adsabs.harvard.edu/abs/2016ApJ...818..112M} {818, 112}

\bibitem[\protect\citeauthoryear{{Mo}, {Mao}  \& {White}}{{Mo}
  et~al.}{1998}]{Mo+98}
{Mo} H.~J.,  {Mao} S.,   {White} S.~D.~M.,  1998, \mn@doi [MNRAS]
  {10.1046/j.1365-8711.1998.01227.x}, \href
  {http://adsabs.harvard.edu/abs/1998MNRAS.295..319M} {295, 319}

\bibitem[\protect\citeauthoryear{{Mulchaey} \& {Jeltema}}{{Mulchaey} \&
  {Jeltema}}{2010}]{Mulchaey+10}
{Mulchaey} J.~S.,  {Jeltema} T.~E.,  2010, \mn@doi [ApJL]
  {10.1088/2041-8205/715/1/L1}, \href
  {http://adsabs.harvard.edu/abs/2010ApJ...715L...1M} {715, L1}

\bibitem[\protect\citeauthoryear{{Muratov}, {Kere{\v s}},
  {Faucher-Gigu{\`e}re}, {Hopkins}, {Quataert}  \& {Murray}}{{Muratov}
  et~al.}{2015}]{Muratov+15}
{Muratov} A.~L.,  {Kere{\v s}} D.,  {Faucher-Gigu{\`e}re} C.-A.,  {Hopkins}
  P.~F.,  {Quataert} E.,   {Murray} N.,  2015, \mn@doi [MNRAS]
  {10.1093/mnras/stv2126}, 454, 2691

\bibitem[\protect\citeauthoryear{{Nahar} \& {Pradhan}}{{Nahar} \&
  {Pradhan}}{2003}]{NaharPradhan03}
{Nahar} S.~N.,  {Pradhan} A.~K.,  2003, \mn@doi [ApJs] {10.1086/377580}, \href
  {http://adsabs.harvard.edu/abs/2003ApJS..149..239N} {149, 239}

\bibitem[\protect\citeauthoryear{{Navarro}, {Frenk}  \& {White}}{{Navarro}
  et~al.}{1997}]{NFW}
{Navarro} J.~F.,  {Frenk} C.~S.,   {White} S.~D.~M.,  1997, ApJ, 490, 493

\bibitem[\protect\citeauthoryear{{Nelson}, {Vogelsberger}, {Genel}, {Sijacki},
  {Kere{\v s}}, {Springel}  \& {Hernquist}}{{Nelson} et~al.}{2013}]{Nelson+13}
{Nelson} D.,  {Vogelsberger} M.,  {Genel} S.,  {Sijacki} D.,  {Kere{\v s}} D.,
  {Springel} V.,   {Hernquist} L.,  2013, \mn@doi [MNRAS]
  {10.1093/mnras/sts595}, 429, 3353

\bibitem[\protect\citeauthoryear{{Nelson}, {Genel}, {Vogelsberger}, {Springel},
  {Sijacki}, {Torrey}  \& {Hernquist}}{{Nelson} et~al.}{2015}]{Nelson+15}
{Nelson} D.,  {Genel} S.,  {Vogelsberger} M.,  {Springel} V.,  {Sijacki} D.,
  {Torrey} P.,   {Hernquist} L.,  2015, \mn@doi [MNRAS] {10.1093/mnras/stv017},
  \href {http://adsabs.harvard.edu/abs/2015MNRAS.448...59N} {448, 59}

\bibitem[\protect\citeauthoryear{{O'Sullivan}, {Forbes}  \&
  {Ponman}}{{O'Sullivan} et~al.}{2001}]{OSullivan+01}
{O'Sullivan} E.,  {Forbes} D.~A.,   {Ponman} T.~J.,  2001, \mn@doi [MNRAS]
  {10.1046/j.1365-8711.2001.04890.x}, \href
  {http://adsabs.harvard.edu/abs/2001MNRAS.328..461O} {328, 461}

\bibitem[\protect\citeauthoryear{{Oppenheimer}, {Dav{\'e}}, {Kere{\v s}},
  {Fardal}, {Katz}, {Kollmeier}  \& {Weinberg}}{{Oppenheimer}
  et~al.}{2010}]{Oppenheimer+10}
{Oppenheimer} B.~D.,  {Dav{\'e}} R.,  {Kere{\v s}} D.,  {Fardal} M.,  {Katz}
  N.,  {Kollmeier} J.~A.,   {Weinberg} D.~H.,  2010, \mn@doi [MNRAS]
  {10.1111/j.1365-2966.2010.16872.x}, \href
  {http://adsabs.harvard.edu/abs/2010MNRAS.406.2325O} {406, 2325}

\bibitem[\protect\citeauthoryear{{Oppenheimer} et~al.,}{{Oppenheimer}
  et~al.}{2016}]{Oppenheimer+16}
{Oppenheimer} B.~D.,  et~al., 2016, preprint, \href
  {http://adsabs.harvard.edu/abs/2016arXiv160305984O} {} (\mn@eprint {arXiv}
  {1603.05984})

\bibitem[\protect\citeauthoryear{{Parrish}, {Quataert}  \& {Sharma}}{{Parrish}
  et~al.}{2010}]{Parrish+10}
{Parrish} I.~J.,  {Quataert} E.,   {Sharma} P.,  2010, \mn@doi [ApJL]
  {10.1088/2041-8205/712/2/L194}, \href
  {http://adsabs.harvard.edu/abs/2010ApJ...712L.194P} {712, L194}

\bibitem[\protect\citeauthoryear{{Parrish}, {McCourt}, {Quataert}  \&
  {Sharma}}{{Parrish} et~al.}{2012}]{Parrish+12}
{Parrish} I.~J.,  {McCourt} M.,  {Quataert} E.,   {Sharma} P.,  2012, \mn@doi
  [MNRAS] {10.1111/j.1365-2966.2012.20650.x}, \href
  {http://adsabs.harvard.edu/abs/2012MNRAS.422..704P} {422, 704}

\bibitem[\protect\citeauthoryear{{Prochaska} \& {Hennawi}}{{Prochaska} \&
  {Hennawi}}{2009}]{ProchaskaHennawi09}
{Prochaska} J.~X.,  {Hennawi} J.~F.,  2009, \mn@doi [ApJ]
  {10.1088/0004-637X/690/2/1558}, \href
  {http://adsabs.harvard.edu/abs/2009ApJ...690.1558P} {690, 1558}

\bibitem[\protect\citeauthoryear{{Prochaska}, {Weiner}, {Chen}, {Mulchaey}  \&
  {Cooksey}}{{Prochaska} et~al.}{2011}]{Prochaska+11}
{Prochaska} J.~X.,  {Weiner} B.,  {Chen} H.-W.,  {Mulchaey} J.,   {Cooksey} K.,
   2011, \mn@doi [ApJ] {10.1088/0004-637X/740/2/91}, \href
  {http://adsabs.harvard.edu/abs/2011ApJ...740...91P} {740, 91}

\bibitem[\protect\citeauthoryear{{Quataert}}{{Quataert}}{2008}]{Quataert08}
{Quataert} E.,  2008, \mn@doi [ApJ] {10.1086/525248}, \href
  {http://adsabs.harvard.edu/abs/2008ApJ...673..758Q} {673, 758}

\bibitem[\protect\citeauthoryear{{Rahmati}, {Pawlik}, {Raicevic}  \&
  {Schaye}}{{Rahmati} et~al.}{2013}]{Rahmati+13}
{Rahmati} A.,  {Pawlik} A.~H.,  {Raicevic} M.,   {Schaye} J.,  2013, \mn@doi
  [MNRAS] {10.1093/mnras/stt066}, \href
  {http://adsabs.harvard.edu/abs/2013MNRAS.430.2427R} {430, 2427}

\bibitem[\protect\citeauthoryear{{Rahmati}, {Schaye}, {Bower}, {Crain},
  {Furlong}, {Schaller}  \& {Theuns}}{{Rahmati} et~al.}{2015}]{Rahmati+15}
{Rahmati} A.,  {Schaye} J.,  {Bower} R.~G.,  {Crain} R.~A.,  {Furlong} M.,
  {Schaller} M.,   {Theuns} T.,  2015, \mn@doi [MNRAS] {10.1093/mnras/stv1414},
  \href {http://adsabs.harvard.edu/abs/2015MNRAS.452.2034R} {452, 2034}

\bibitem[\protect\citeauthoryear{{Rees} \& {Ostriker}}{{Rees} \&
  {Ostriker}}{1977}]{ReesOstriker+77}
{Rees} M.~J.,  {Ostriker} J.~P.,  1977, MNRAS, \href
  {http://adsabs.harvard.edu/abs/1977MNRAS.179..541R} {179, 541}

\bibitem[\protect\citeauthoryear{{Rubin}, {Prochaska}, {M{\'e}nard}, {Murray},
  {Kasen}, {Koo}  \& {Phillips}}{{Rubin} et~al.}{2011}]{Rubin+11}
{Rubin} K.~H.~R.,  {Prochaska} J.~X.,  {M{\'e}nard} B.,  {Murray} N.,  {Kasen}
  D.,  {Koo} D.~C.,   {Phillips} A.~C.,  2011, \mn@doi [ApJ]
  {10.1088/0004-637X/728/1/55}, \href
  {http://adsabs.harvard.edu/abs/2011ApJ...728...55R} {728, 55}

\bibitem[\protect\citeauthoryear{{Rudie} et~al.,}{{Rudie}
  et~al.}{2012}]{Rudie+12}
{Rudie} G.~C.,  et~al., 2012, \mn@doi [ApJ] {10.1088/0004-637X/750/1/67}, \href
  {http://adsabs.harvard.edu/abs/2012ApJ...750...67R} {750, 67}

\bibitem[\protect\citeauthoryear{{Ruszkowski} \& {Oh}}{{Ruszkowski} \&
  {Oh}}{2010}]{RuszkowskiOh10}
{Ruszkowski} M.,  {Oh} S.~P.,  2010, \mn@doi [ApJ]
  {10.1088/0004-637X/713/2/1332}, \href
  {http://adsabs.harvard.edu/abs/2010ApJ...713.1332R} {713, 1332}

\bibitem[\protect\citeauthoryear{{Salem}, {Bryan}  \& {Corlies}}{{Salem}
  et~al.}{2016}]{Salem+16}
{Salem} M.,  {Bryan} G.~L.,   {Corlies} L.,  2016, \mn@doi [MNRAS]
  {10.1093/mnras/stv2641}, \href
  {http://adsabs.harvard.edu/abs/2016MNRAS.456..582S} {456, 582}

\bibitem[\protect\citeauthoryear{Sharma, McCourt, Quataert  \& Parrish}{Sharma
  et~al.}{2012a}]{Sharma+12a}
Sharma P.,  McCourt M.,  Quataert E.,   Parrish I.~J.,  2012a, Monthly Notices
  of the Royal Astronomical Society, 420, 3174

\bibitem[\protect\citeauthoryear{Sharma, McCourt, Parrish  \& Quataert}{Sharma
  et~al.}{2012b}]{Sharma+12b}
Sharma P.,  McCourt M.,  Parrish I.~J.,   Quataert E.,  2012b, Monthly Notices
  of the Royal Astronomical Society, 427, 1219

\bibitem[\protect\citeauthoryear{{Shen}, {Madau}, {Guedes}, {Mayer},
  {Prochaska}  \& {Wadsley}}{{Shen} et~al.}{2013}]{Shen+13}
{Shen} S.,  {Madau} P.,  {Guedes} J.,  {Mayer} L.,  {Prochaska} J.~X.,
  {Wadsley} J.,  2013, \mn@doi [ApJ] {10.1088/0004-637X/765/2/89}, \href
  {http://adsabs.harvard.edu/abs/2013ApJ...765...89S} {765, 89}

\bibitem[\protect\citeauthoryear{{Silk}}{{Silk}}{1977}]{Silk77}
{Silk} J.,  1977, \mn@doi [ApJ] {10.1086/154972}, \href
  {http://adsabs.harvard.edu/abs/1977ApJ...211..638S} {211, 638}

\bibitem[\protect\citeauthoryear{{Springel} et~al.,}{{Springel}
  et~al.}{2005}]{Millennium}
{Springel} V.,  et~al., 2005, \mn@doi [Nature] {10.1038/nature03597}, 435, 629

\bibitem[\protect\citeauthoryear{{Steidel}, {Erb}, {Shapley}, {Pettini},
  {Reddy}, {Bogosavljevi{\'c}}, {Rudie}  \& {Rakic}}{{Steidel}
  et~al.}{2010}]{Steidel+10}
{Steidel} C.~C.,  {Erb} D.~K.,  {Shapley} A.~E.,  {Pettini} M.,  {Reddy} N.,
  {Bogosavljevi{\'c}} M.,  {Rudie} G.~C.,   {Rakic} O.,  2010, \mn@doi [ApJ]
  {10.1088/0004-637X/717/1/289}, \href
  {http://adsabs.harvard.edu/abs/2010ApJ...717..289S} {717, 289}

\bibitem[\protect\citeauthoryear{{Stern}, {Hennawi}, {Prochaska}  \&
  {Werk}}{{Stern} et~al.}{2016}]{Stern+16}
{Stern} J.,  {Hennawi} J.~F.,  {Prochaska} J.~X.,   {Werk} J.~K.,  2016,
  preprint, \href {http://adsabs.harvard.edu/abs/2016arXiv160402168S} {}
  (\mn@eprint {arXiv} {1604.02168})

\bibitem[\protect\citeauthoryear{{Stocke}, {Keeney}, {Danforth}, {Shull},
  {Froning}, {Green}, {Penton}  \& {Savage}}{{Stocke} et~al.}{2013}]{Stocke+13}
{Stocke} J.~T.,  {Keeney} B.~A.,  {Danforth} C.~W.,  {Shull} J.~M.,  {Froning}
  C.~S.,  {Green} J.~C.,  {Penton} S.~V.,   {Savage} B.~D.,  2013, \mn@doi
  [ApJ] {10.1088/0004-637X/763/2/148}, \href
  {http://adsabs.harvard.edu/abs/2013ApJ...763..148S} {763, 148}

\bibitem[\protect\citeauthoryear{{Stone}, {Gardiner}, {Teuben}, {Hawley}  \&
  {Simon}}{{Stone} et~al.}{2008}]{Stone+08}
{Stone} J.~M.,  {Gardiner} T.~A.,  {Teuben} P.,  {Hawley} J.~F.,   {Simon}
  J.~B.,  2008, \mn@doi [ApJs] {10.1086/588755}, 178, 137

\bibitem[\protect\citeauthoryear{{Suresh}, {Rubin}, {Kannan}, {Werk},
  {Hernquist}  \& {Vogelsberger}}{{Suresh} et~al.}{2015a}]{Suresh+15b}
{Suresh} J.,  {Rubin} K.~H.~R.,  {Kannan} R.,  {Werk} J.~K.,  {Hernquist} L.,
  {Vogelsberger} M.,  2015a, preprint, \href
  {http://adsabs.harvard.edu/abs/2015arXiv151100687S} {} (\mn@eprint {arXiv}
  {1511.00687})

\bibitem[\protect\citeauthoryear{{Suresh}, {Bird}, {Vogelsberger}, {Genel},
  {Torrey}, {Sijacki}, {Springel}  \& {Hernquist}}{{Suresh}
  et~al.}{2015b}]{Suresh+15a}
{Suresh} J.,  {Bird} S.,  {Vogelsberger} M.,  {Genel} S.,  {Torrey} P.,
  {Sijacki} D.,  {Springel} V.,   {Hernquist} L.,  2015b, \mn@doi [MNRAS]
  {10.1093/mnras/stu2762}, \href
  {http://adsabs.harvard.edu/abs/2015MNRAS.448..895S} {448, 895}

\bibitem[\protect\citeauthoryear{{Sutherland} \& {Dopita}}{{Sutherland} \&
  {Dopita}}{1993}]{SD93}
{Sutherland} R.~S.,  {Dopita} M.~A.,  1993, \mn@doi [ApJs] {10.1086/191823},
  88, 253

\bibitem[\protect\citeauthoryear{{Tepper-Garc{\'{\i}}a}, {Richter}, {Schaye},
  {Booth}, {Dalla Vecchia}, {Theuns}  \& {Wiersma}}{{Tepper-Garc{\'{\i}}a}
  et~al.}{2011}]{TepperGarcia+11}
{Tepper-Garc{\'{\i}}a} T.,  {Richter} P.,  {Schaye} J.,  {Booth} C.~M.,  {Dalla
  Vecchia} C.,  {Theuns} T.,   {Wiersma} R.~P.~C.,  2011, \mn@doi [MNRAS]
  {10.1111/j.1365-2966.2010.18123.x}, \href
  {http://adsabs.harvard.edu/abs/2011MNRAS.413..190T} {413, 190}

\bibitem[\protect\citeauthoryear{{Thom} et~al.,}{{Thom} et~al.}{2012}]{Thom+12}
{Thom} C.,  et~al., 2012, \mn@doi [ApJL] {10.1088/2041-8205/758/2/L41}, \href
  {http://adsabs.harvard.edu/abs/2012ApJ...758L..41T} {758, L41}

\bibitem[\protect\citeauthoryear{{Thompson}, {Quataert}, {Zhang}  \&
  {Weinberg}}{{Thompson} et~al.}{2016}]{Thompson+16}
{Thompson} T.~A.,  {Quataert} E.,  {Zhang} D.,   {Weinberg} D.~H.,  2016,
  \mn@doi [MNRAS] {10.1093/mnras/stv2428}, \href
  {http://adsabs.harvard.edu/abs/2016MNRAS.455.1830T} {455, 1830}

\bibitem[\protect\citeauthoryear{Tumlinson et~al.,}{Tumlinson
  et~al.}{2011}]{Tumlinson+11}
Tumlinson J.,  et~al., 2011, Science, 334, 948

\bibitem[\protect\citeauthoryear{{Tumlinson} et~al.,}{{Tumlinson}
  et~al.}{2013}]{Tumlinson+13}
{Tumlinson} J.,  et~al., 2013, \mn@doi [ApJ] {10.1088/0004-637X/777/1/59},
  \href {http://adsabs.harvard.edu/abs/2013ApJ...777...59T} {777, 59}

\bibitem[\protect\citeauthoryear{{Turk}, {Smith}, {Oishi}, {Skory}, {Skillman},
  {Abel}  \& {Norman}}{{Turk} et~al.}{2011}]{yt}
{Turk} M.~J.,  {Smith} B.~D.,  {Oishi} J.~S.,  {Skory} S.,  {Skillman} S.~W.,
  {Abel} T.,   {Norman} M.~L.,  2011, \mn@doi [ApJs]
  {10.1088/0067-0049/192/1/9}, \href
  {http://adsabs.harvard.edu/abs/2011ApJS..192....9T} {192, 9}

\bibitem[\protect\citeauthoryear{{Veilleux}, {Cecil}  \&
  {Bland-Hawthorn}}{{Veilleux} et~al.}{2005}]{Veilleux+2005}
{Veilleux} S.,  {Cecil} G.,   {Bland-Hawthorn} J.,  2005, \mn@doi [AR\&A]
  {10.1146/annurev.astro.43.072103.150610}, \href
  {http://adsabs.harvard.edu/abs/2005ARA%26A..43..769V} {43, 769}

\bibitem[\protect\citeauthoryear{Voit \& Donahue}{Voit \&
  Donahue}{2015}]{Voit+15}
Voit G.~M.,  Donahue M.,  2015, The Astrophysical Journal Letters, 799, L1

\bibitem[\protect\citeauthoryear{{Voit}, {Bryan}, {O'Shea}  \&
  {Donahue}}{{Voit} et~al.}{2015}]{Voit+15b}
{Voit} G.~M.,  {Bryan} G.~L.,  {O'Shea} B.~W.,   {Donahue} M.,  2015, \mn@doi
  [ApJL] {10.1088/2041-8205/808/1/L30}, \href
  {http://adsabs.harvard.edu/abs/2015ApJ...808L..30V} {808, L30}

\bibitem[\protect\citeauthoryear{{Weiner} et~al.,}{{Weiner}
  et~al.}{2009}]{Weiner+09}
{Weiner} B.~J.,  et~al., 2009, \mn@doi [ApJ] {10.1088/0004-637X/692/1/187},
  \href {http://adsabs.harvard.edu/abs/2009ApJ...692..187W} {692, 187}

\bibitem[\protect\citeauthoryear{{Werk} et~al.,}{{Werk} et~al.}{2014}]{Werk+14}
{Werk} J.~K.,  et~al., 2014, \mn@doi [ApJ] {10.1088/0004-637X/792/1/8}, 792, 8

\bibitem[\protect\citeauthoryear{{Werk} et~al.,}{{Werk} et~al.}{2016}]{Werk+16}
{Werk} J.~K.,  et~al., 2016, preprint, \href
  {http://adsabs.harvard.edu/abs/2016arXiv160900012W} {} (\mn@eprint {arXiv}
  {1609.00012})

\bibitem[\protect\citeauthoryear{{Wiersma}, {Schaye}  \& {Smith}}{{Wiersma}
  et~al.}{2009}]{Wiersma+09}
{Wiersma} R.~P.~C.,  {Schaye} J.,   {Smith} B.~D.,  2009, \mn@doi [MNRAS]
  {10.1111/j.1365-2966.2008.14191.x}, \href
  {http://adsabs.harvard.edu/abs/2009MNRAS.393...99W} {393, 99}

\bibitem[\protect\citeauthoryear{{Yang}, {Mo}  \& {van den Bosch}}{{Yang}
  et~al.}{2009}]{Yang+09}
{Yang} X.,  {Mo} H.~J.,   {van den Bosch} F.~C.,  2009, \mn@doi [ApJ]
  {10.1088/0004-637X/695/2/900}, \href
  {http://adsabs.harvard.edu/abs/2009ApJ...695..900Y} {695, 900}

\bibitem[\protect\citeauthoryear{{van de Voort} \& {Schaye}}{{van de Voort} \&
  {Schaye}}{2012}]{vandeVoort+12}
{van de Voort} F.,  {Schaye} J.,  2012, \mn@doi [MNRAS]
  {10.1111/j.1365-2966.2012.20949.x}, \href
  {http://adsabs.harvard.edu/abs/2012MNRAS.423.2991V} {423, 2991}

\bibitem[\protect\citeauthoryear{{van de Voort}, {Quataert}, {Hopkins},
  {Faucher-Gigu{\`e}re}, {Feldmann}, {Kere{\v s}}, {Chan}  \& {Hafen}}{{van de
  Voort} et~al.}{2016}]{vandeVoort+16}
{van de Voort} F.,  {Quataert} E.,  {Hopkins} P.~F.,  {Faucher-Gigu{\`e}re}
  C.-A.,  {Feldmann} R.,  {Kere{\v s}} D.,  {Chan} T.~K.,   {Hafen} Z.~H.,
  2016, preprint, \href {http://adsabs.harvard.edu/abs/2016arXiv160401397V} {}
  (\mn@eprint {arXiv} {1604.01397})

\makeatother
\end{thebibliography}

\label{lastpage}

\end{document}